# Atomic-scale Electronic Structure of the Cuprate Pair Density Wave State Coexisting with Superconductivity

Peayush Choubey[1,2§,] Sang Hyun Joo[3§], K. Fujita[4], Zengyi Du[4], S. D. Edkins[5],
M. H. Hamidian[6], H. Eisaki[7], S. Uchida[7], A. P. Mackenzie[8], Jinho Lee[3],
J.C. Séamus Davis[9,10,11] and P.J. Hirschfeld[12]

1. *Institut für Theoretische Physik III, Ruhr-Universität Bochum, D-44801 Bochum, Germany.*
2. *Department of Physics, Indian Institute of Science, Bengaluru-560012, India.*
3. *Department of Physics and Astron., Seoul National University, Seoul 08826, Korea.*
4. *CMPMS Department, Brookhaven National Laboratory, Upton, NY 11973, USA.*
5. *Department of Applied Physics, Stanford University, Stanford, CA 94305*
6. *Department of Physics, Harvard University, Cambridge MA, USA.*
7. *Inst. of Advanced Industrial Science and Tech., Tsukuba, Ibaraki 305-8568, Japan.*
8. *Max-Planck Institute for Chemical Physics of Solids, D-01187 Dresden, Germany.*
9. *LASSP, Department of Physics, Cornell University, Ithaca NY 14850, USA*
10. *Clarendon Laboratory, University of Oxford, Oxford, OX1 3PU, UK*
11. *Department of Physics, University College Cork, Cork T12 R5C, Ireland*
12. *Department of Physics, University of Florida, Gainesville, FL, USA*
§ *These authors contributed equally to this project.*

ABSTRACT**:** The defining characteristic of hole-doped cuprates is $d$-wave high temperature superconductivity. However, intense theoretical interest is now focused on whether a pair density wave state (PDW) could coexist with cuprate superconductivity (D. F. Agterberg *et al., Annual Review of Condensed Matter Physics 11, 231 (2020)*). Here, we use a strong-coupling mean-field theory of cuprates, to model the atomic-scale electronic structure of an eight-unit-cell periodic, $d$-symmetry form factor, pair density wave (PDW) state coexisting with $d$-wave superconductivity (DSC). From this PDW+DSC model, the atomically-resolved density of Bogoliubov quasiparticle states $N(r, E)$ is predicted at the terminal BiO surface of $Bi_2Sr_2CaCu_2O_8$ and compared with high-precision electronic visualization experiments using spectroscopic imaging STM. The PDW+DSC model predictions include the intra-unit-cell structure and periodic modulations of $N(r, E)$, the modulations of the coherence peak energy $\Delta_p(r)$, and the characteristics of Bogoliubov quasiparticle interference in scattering-wavevector space ($q$ −space). Consistency between all these predictions and the corresponding experiments indicates that lightly hole-doped $Bi_2Sr_2CaCu_2O_8$ does contain a PDW+DSC state. Moreover, in the model the PDW+DSC state becomes unstable to a pure DSC state at a critical hole density $p$*, with empirically equivalent phenomena occurring in the experiments. All these results are consistent with a picture in which the cuprate translational symmetry breaking state is a PDW, the observed charge modulations are its consequence, the antinodal pseudogap is that of the PDW state, and the cuprate critical point at $p^* \approx 19\%$ occurs due to disappearance of this PDW.

*Keywords:* Cuprate Pseudogap, Pair Density Wave State, Quasiparticle Interference,

*The Pair Density Wave and Pseudogap States of Cuprates*

**1**      In the elementary undoped $CuO_2$ plane, each Cu $d_{x^2-y^2}$ orbital is occupied by a single electron and, because the energy required to doubly occupy this orbital is $U \sim 3 \, eV$, a Mott insulator (MI) state develops[1,2]. The superexchange spin-spin interaction energy between neighboring $d_{x^2-y^2}$ electrons is $J \sim 150 \, meV$, leading to a robust antiferromagnetic (AF) phase[3,4] (Fig 1a). But this AF insulating state vanishes with the removal of as little as 3% of the electrons per Cu site (hole-density $p$=3%), to reveal the pseudogap (PG) state in a region of the phase diagram bounded by $p < p^*$ and temperatures $T < T^*(p)$ (Fig. 1a). Key characteristics of the PG state include[3,4] a steep drop in both magnetic susceptibility and $c$-axis conductivity; an apparently incomplete Fermi surface consisting of coherent quasiparticle states on four $k-$space arcs neighboring $k \approx (\pm \pi/2a, \pm \pi/2a)$ ; an energy gap $\Delta^*$ in the spectrum of quasiparticle states near $k \approx (\pm \pi/a, 0); (0, \pm \pi/a)$; and the depletion of the average density of electronic states $N(E)$ for $|E| < \Delta^*$ where $\Delta^*(p)$ diminishes to zero at $p = p*$ (Fig. 1a). A mean-field energy-gap in the spectrum of coherent $k$-space quasiparticles occurring only near $k \approx (\pm \pi/a, 0); (0, \pm \pi/a, 0)$ could provide a simple phenomenological explanation for virtually all these PG characteristics, but no comprehensive microscopic theory for the PG phase has yet been established.

**2**      Extensive evidence has recently emerged for electronic symmetry breaking within the PG phase (Fig. 1a). Bulk probes of charge density find translational symmetry breaking in a density-wave (DW) state with axial wavevectors $Q = (Q, 0); (0, Q)$ parallel to the $CuO_2$ axes[1,2,5]. Similarly, direct visualization with sub-unit-cell resolution using single-electron tunneling in $Bi_2Sr_2CaCu_2O_8$ and $Ca_{2-x}Na_xCuO_2Cl_2$ reveals intense electronic structure modulations[6,7] that are locally unidirectional[7, 8], exhibit lattice-commensurate periodicity[9,10] for all $p < p*$ [11], have a $d$-symmetry form factor[8, 12], and are concentrated at particle-hole symmetric energies[13] $|E| \approx \Delta^*(p)$. The spatial configurations consist of nanoscale regions within which the modulations are commensurate and unidirectional along either[7,8,10,12] $(Q, 0)$ or $(0, Q)$. A complete theoretical explanation for the microscopic origin of these complex atomic-scale electronic structures[6,7,8,9,10,12] has never been elucidated. Figure 1b shows a representative example of $Z(r, E = \Delta) = N(r, +\Delta) / N(r, -\Delta)$ for this state at $p \approx$ 8%, along with the simultaneously measured topograph $T(r)$ at the BiO layer with the Cu sites indicated by crosses. Figure 1c shows the measured differential tunneling conductance g($E$) averaged over the whole FOV of Fig. 1b, and identifies the two characteristic energies $\Delta_1$ and $\Delta_0$. Thus, a key challenge for cuprate studies is to identify microscopically the broken symmetry state in Fig. 1b,c that coexists with the DSC phase and to determine its relationship to the pseudogap.

**3**     Because the strong electron-electron interactions subtending the MI state persist even when the long-range AF order disappears (Fig. 1a), strong-coupling theory also seems necessary in the PG phase. One frequently recurring consequence[14- 22] of such theories is the existence of a state that breaks translational symmetry by modulating the electron-pairing field as

$$\Delta_1 (r) = F_P \Delta_1 \left[ e^{iQ_P \cdot r} + e^{-iQ_P \cdot r} \right] \qquad (1)$$

This is a pair density wave state for which $\Delta_1$ is the magnitude of the PDW order parameter and $F_P$ is its form-factor symmetry. Moreover, such a strong-coupling PDW state intertwines[23] the modulations of electron-pair field, of the site/bond charge density, and of spin density. Focus on whether such a PDW state exists in the ground state of cuprates has been further motivated by growing experimental evidence[24- 29,30] that is consistent therewith. Obviously, if this PDW state occurs, it must coexist in the $T \to 0$ ground state together with the robust $d$-wave superconductivity (Fig. 1a). Therefore, an urgent research priority is to understand the atomic-scale electronic structure of a PDW coexisting with a DSC state. This is quite challenging because it requires a theoretical description of PDW+DSC electronic structure at the intra-unit-cell scale in $r$-space, and simultaneously throughout a Brillouin zone in $k$-space that is strongly altered by the PDW's existence.

*Theory for Cuprate Pair Density Wave and coexisting Superconductivity*

**4**     A classic theory for hole-doped CuO₂ plane electronic structure is based on the *t-J* model, in which electrons hop with matrix element $t$ between Cu $d_{x^2-y^2}$ orbitals, onsite Coulomb energy $U \to \infty$ to completely prevent their double occupancy, resulting in strong antiferromagnetic exchange interactions *J=4t²/U*. Its Hamiltonian is $= -\sum_{(i,j),\sigma} P_G t_{ij}(c_{i\sigma}^\dagger c_{j\sigma} + h.c.) P_G + J \sum_{<i,j>} S_i \cdot S_j$ , where the operator $P_G$ projects out all doubly occupied orbitals from the Hilbert space (SI Appendix, Section A). A renormalized mean-field theory (RMFT) approximation to this *t-J* model is then of great utility in describing the CuO₂ plane physics[31]; it replaces the exact projection $P_G$ operation with renormalization factors $g_i^t$ and $g_i^s$ determine by the average number of charge and spin configurations permissible at every Cu site. The resulting Hamiltonian can be decoupled into a simpler but diagonalizable approximation by using the mean-fields describing on-site hole density $\delta_i$ , bond field $\chi_{ij\sigma}$ , and electron-pair field $\Delta_{ij\sigma}$ (SI Appendix, Section A). Subsequent variational minimization of the ground state energy with respect to the unprojected wavefunction $|\Psi_0 >$ leads to a set of Bogoliubov-de Gennes (BdG) equations, together with self-consistency conditions on the mean fields. To allow breaking of translational symmetry within RMFT, site-specific and bond-specific renormalization factors $g_{i,j}^t$ and $g_{i,j}^s$ for charge and spin are introduced[17]. To obtain a PDW+ DSC solution, the BdG equations are initialized with a set of order parameter fields modulating at wavevector $Q_P = (1/8,0)2\pi/a_0$, and a self-consistent solution is found. From the consequent many-body wavefunction $\Psi_0(r)$ of this broken-symmetry state, the net charge on each Cu site

$$\delta_i = 1 - < \Psi_0 |\sum_\sigma n_{i\sigma}|\Psi_0 > \qquad (2)$$

the bond field between adjacent sites *i,j*

$$\chi_{ij\sigma} = < \Psi_0 |c_{i\sigma}^\dagger c_{j\sigma}|\Psi_0 > \qquad (3)$$

and the electron-pair field on the bond between adjacent sites *i,j*

$$\Delta_{ij\sigma} = \sigma < \Psi_0 |c_{i\sigma} c_{j\bar{\sigma}}|\Psi_0 > \qquad (4)$$

can all be calculated. We note that the commensurate PDW+DSC studied here is not the ground state of the cuprate RMFT Hamiltonian, but that its energy above the homogeneous ground state is so tiny[17,20,21] (∼1meV per Cu as discussed in SI Appendix, Section A) that it may be stabilized by a variety of means, including disorder.

*Comparison of Pair Density Wave plus Superconductivity Theory with Experiment*

*5*      Using this model, we explore the atomic-scale characteristics of a unidirectional, $\lambda =$ $8a_0$ PDW state coexisting with uniform DSC. Figure 2a shows the predicted hole density $\delta_i$ on the Cu sites exhibiting dominant $\lambda = 4a_0$ modulations, together with the projected $d$-wave superconducting order parameter amplitude $\Delta_i$ on the same sites. The corresponding Fourier components of the hole density $\delta(q)$ and electron-pair field $\Delta(q)$ are shown in Fig. 2b. Here it is important to note that the PDW-induced charge modulation amplitude at $\lambda = 8a_0$ is extremely weak for our standard set of parameters (see caption of Figure 2), whereas the induced charge modulation amplitude at $\lambda = 4a_0$ is dominant. Next, in Fig. 2c, we exhibit the projection of the $8a_0$ periodic charge modulations onto the three symmetry-allowed channels *s, s'* and *d,* utilizing the same definitions as, e.g. in Ref. [12] (see SI Appendix, Section A). The charge distribution over the Cu and O sites is clearly characterized by a pronounced $d$-symmetry form factor (dFF), as has been observed directly in experiments[8,12].

*6*      For comparison with underdoped $Bi_2Sr_2CaCu_2O_8$ measurements, we next use the RMFT PDW+DSC model and evaluate its quasiparticle states with intra-unit-cell resolution using a Wannier function-based method[21,22]. This is designed to allow quantitative predictions of electronic structure in $r$-space, $q$-space and $k$-space of a PDW+DSC state. The band structure parameterization is $t$=400meV, $t'$=-0.3t and $J$=0.3t, all representing $Bi_2Sr_2CaCu_2O_8$ at $p \approx 0.08$. For this parameter set, we calculate the quasiparticle Greens functions in the self-consistently obtained PDW+DSC state, with the unidirectional PDW wavevector $Q_P = (1/8,0)2\pi/a_0$ that is modulating parallel to the $x$-axis but lattice-periodic along the y-axis, and find the PDW spectral gap $\Delta_1 \approx 0.3t \approx 100meV$, and gap associated with uniform DSC $\Delta_0 \approx 0.07t \approx 25meV$ (SI Appendix, Section A). Moreover, while all previous RMFT studies of cuprates yield only the Cu Site-specific Greens function matrix $G_{ij\sigma}(E)$ within the $CuO_2$ plane, the experimental measurements of electron-tunneling probability are actually made at a continuum of locations just above the crystal termination

BiO layer of $Bi_2Sr_2CaCu_2O_8$ (Fig. 1b). Therefore, we use first-principles Cu $d_{x^2-y^2}$ Wannier functions $W_i(r)$ to make quantitative predictions of the $r$-space Green's functions $G_\sigma(r, E) = \sum_{ij} G_{ij\sigma}(E)W_i(r)W_j^*(r)$ of a PDW+DSC state, everywhere at a height 0.4nm above BiO terminal plane[32,33] (SI Appendix, Section B). We emphasize that none of the mean fields $\delta_i$, $\chi_{ij\sigma}$ and $\Delta_{ij\sigma}$ are related simply to the local quasiparticle density of states $N(r, E) = \sum_\sigma -\frac{1}{\pi} \text{Im } G_\sigma(r, E)$, which must instead be determined from the Bogoliubov quasiparticle eigenstates[20,21,24,25,28,34,35,36] that enter the lattice Green's function $G_{ij\sigma}$. Fig. 2d shows the theoretical $N(\mathbf{r}, E)$ for the DSC+PDW state at various points in the modulated state, as identified in the inset. Note that a shoulder-like feature is present at nearly identical particle-hole symmetric energies $\pm\Delta_0$ for all Cu sites, such that we associate it with the uniform $d$-wave superconducting condensate DSC. By contrast, this model predicts that the energy $\Delta_p$ at which the "coherence peak" occurs ($\omega > 0$) varies from atom to atom in real space; its largest magnitude may be associated with the PDW amplitude $\Delta_1$.

7  The bias dependence of $Z(r, E) = N(r, +E) / N(r, -E)$ from our PDW+DSC model are then predicted for comparison with experiment. Figure 3a-f shows these $Z(r, E)$ data focusing on the energy range $0.5\Delta_1 \lesssim E \lesssim 1.5\Delta_1$, within which they exhibit a comprehensive $d$-symmetry form factor (SI Appendix, Section B). This effect can be seen directly because the model $Z(r, E)$ has intra-unit-cell precision. Consider the three sublattices making up the primary features of the $Z(r, E)$ image: $Cu(r, E)$ containing only $Z(r, E)$ at copper sites and $O_x(r, E)$ and $O_y(r, E)$, containing only $Z(r, E)$ at the $x/y$-axis oxygen sites. By definition, in a $d$-symmetry form factor charge density wave, modulations on the $O_x(r, E)$ and $O_y(r, E)$ sites are out of phase by $\pi$. In our PDW+DSC model, such phenomena occur at both $Q_P$ and $2Q_P$, first appear near $E \approx \Delta_1/2$, are intense surrounding $E \approx \Delta_1$ and eventually disappear near $E \approx 2\Delta_1$ (SI Appendix, Section B and Ref. 8). This energy dependence, including both the dominance of the dFF in a wide range of energies surrounding $\Delta_1$, as well as the low-energy dominance of the $s'$ form factor, were demonstrated earlier using the same theory[21] but for an incommensurate wavevector.

**8**      For experimental comparison with our PDW+DSC model, we visualize the electronic structure[37] using spectroscopic imaging STM (SISTM) measurements of STM-tip-sample differential electron tunneling conductance $dI/dV(r,V) \equiv g(r,V)$. We study the terminal BiO layer of $Bi_2Sr_2CaCu_2O_8$ for a range of tip-sample voltage differences $V$ and at T=4.2K. In theory, $g(r,V) \propto N(r,E)/\int_0^{eV_s} N(r,E)dE$ ($V_s$ is the junction-formation voltage). The ratio $Z(r,V) = g(r,+V)/g(r,-V)$ is widely used in such visualization studies because, even if $\int_0^{eV_s} N(r,E)dE$ is heterogeneous, it yields a valid measure of periodicities and broken symmetries[37]. Figures 3g-l show the measured $Z(r,E)$ in the energy range $0.5\Delta_1 \lesssim E \lesssim 1.5\Delta_1$, with each panel shown side-by-side with the equivalent energy in the model. The most intense modulations occur at $E \approx \Delta_1(p)$ for all $p < 0.19$ (Ref.8, 13), and all through this energy range they exhibit a comprehensive $d$-symmetry form factor[8,12,13]. Therefore, correspondence between theoretical $Z(r,E)$ from the PDW+DSC model (Fig. 3a-f) and the measured $Z(r,E)$ (Fig. 3g-l) is observed to be excellent over a wide energy range. This can be quantified by measuring the cross-correlation value of each pair of theory: experiment $Z\left(r,E/\Delta_1\right)$ images . The result, shown in Fig. 3m, demonstrates strong cross-correlations between theory-experiment pairs of $Z(r,E)$ images throughout the energy range. Therefore, predictions of PDW+DSC theory, on distance scales ranging from $8a_0$ down to sub-unit cell, corresponds strongly and in detail to the complex patterns of quasiparticle states observed in the broken symmetry state of $p<p^*$ $Bi_2Sr_2CaCu_2O_8$.

**9**      Next, as seen in Fig. 2d, the theory predicts that coherence peak energy $\Delta_p$ varies substantially from one unit cell to the next within the $\lambda = 8a_0$ PDW. In Fig. 4a we show the theoretical gap map for the PDW+DSC state obtained by identifying the coherence peak energy $\Delta_p(r)$ for $\omega>0$ at all intra-unit cell points over an area of 8x12 unit cells, while Fig. 4b shows the gap map obtained by using the same algorithm to determine $\Delta_p(r)$ from measured dI/dV spectra. Both theory and experiment show $8a_0$ periodic $\Delta_p(r)$ modulations within which there are smaller atomically resolved variations that exhibit common characteristics but are not identical, most likely because of inadequacies in the DFT-derived Wannier functions in representing underdoped cuprates.

**10**    Finally, we consider the effects of a PDW+DSC state on Bogoliubov quasiparticle scattering interference[38] (BQPI). This occurs when an impurity atom scatters quasiparticles, which then interfere to produce characteristic modulations of $N(r, E)$ surrounding each impurity atom. Local maxima in $Z(q, E)$, the power-spectral-density Fourier transform of $Z(r, E)$, reveal the sets of energy dispersive wavevectors $q_i(E)$ generated by the scattering interference[11]. A BQPI data set thus consists of a sequence of $Z(q, E)$ images spanning the energy range of interest, from which an efficient synopsis over all the QPI modulations can be achieved[11] using $\Lambda(q, \Delta) = \sum_{E=0}^{\Delta} Z(q, E)$ . The key utility here is that $\Lambda(q, \Delta)$ provides an efficient and characteristic "fingerprint" of whatever ordered state(s) controls the $q_i(E)$ of the BQPI processes. For our PDW+DSC model, we calculate $Z(q, E)$ using a simple point-like scatterer within the RMFT framework (SI Appendix, Section C). From these model $Z(q, E)$ images, the predicted $\Lambda(q, \Delta_0) = \sum_{E=0}^{\Delta_0} Z(q, E)$ is determined and shown in Fig. 5a. This contains the overall $\Lambda(q, \Delta_0)$ fingerprint expected of the BQPI in this PDW+DSC state. For comparison, Fig. 5b includes the predicted $\Lambda(q, \Delta_0)$ of a simple $d$-wave superconductor with a Fermi surface at $p$=23%. It reflects the familiar peaks characteristic of dispersing Bogoliubov quasiparticles, including the tracing of the Fermi surface by the peak conventionally labelled $q_4$ as bias voltage is varied[11]. Clearly, the PDW+DSC $\Lambda(q, \Delta_0)$ in Fig 5a is very different, exhibiting very weak dispersion and thus producing sharper $q$-space spots, as well as the absence of any $q_4$ scattering interference near edges of $q$-space reciprocal unit cell. To compare these predictions for $\Lambda(q, \Delta_0)$ with and without the PDW order to experiments, we show measured $\Lambda(q) = \sum_{E \cong 0}^{\Delta_0} Z(q, E)$ for both low $p$ and high $p$ in Fig. 5c,d. The very distinct characteristics of $\Lambda(q, \Delta_0)$ observed at low $p$ and high $p$, are in striking agreement with the PDW+DSC model predictions for $\Lambda(q, \Delta_0)$ with and without the PDW state respectively. Moreover, the empirical BQPI phenomena in Fig. 5c are a characteristic of the pseudogap region of the phase diagram[11] whereas the PDW+DSC model that predicts them (Fig. 5a) does not require a separate pseudogap to be introduced because the $k$-space structure of the PDW is what gaps antinodes[34]. Finally, in our PDW+DSC model, we find that increasing $p$ leads to an instability of the PDW order towards a uniform DSC state via a weak first order transition at $p \approx 0.18$ (SI Appendix, Section A), whereas the experimentally observed disappearance of translational symmetry breaking simultaneous with the reappearance of low energy quasiparticles at the antinodes[11], occurs at $p \approx 0.19$.

*Pair Density Wave plus Superconductivity Theory and Other Techniques*

**11**    We note that NMR and XRAY studies (primarily of $YBa_2Cu_3O_7$ and $La_2BaCuO_4$)[39] appear consistent with SISTM visualizations of incommensurate electronic density modulations (primarily of $Bi_2Sr_2CaCu_2O_8$ and $Ca_{2-x}Na_xCuO_2Cl_2$)[37], and to have a common microscopic cause[40,41]. But in our current study, we have compared carefully to atomic scale studies of $Bi_2Sr_2CaCu_2O_8$ that manifest short-range commensurate periodicity in electronic structure images[7,9,10,12]. These two phenomenologies may appear mutually contradictory. However, the XRAY and Fourier transform STM (FTSTM) studies focus only on a specific wavevector, and this procedure discards all the other information distributed throughout $q$-space. In $Bi_2Sr_2CaCu_2O_8$, that data actually contains the extremely complex information on the effects of form factor symmetry, and on how the commensurate, unidirectional electronic-structure patters proliferate throughout real-space[10]. Thus, the coexistence of a local maximum at incommensurate wavevector in XRAY or FTSTM measurements, with commensurate electronic structures distributed throughout real space, is a demonstrable characteristic of $Bi_2Sr_2CaCu_2O_8$ electronic structure[10]. The most likely resolution is that the measured incommensurate behavior in XRAY or FTSTM experiments represent really *discommensurate* behavior, where commensurate, unidirectional DW are coupled by random phase slips. Indeed, by using advanced techniques, these phase slips are identifiable directly in the N($r$,E) modulations of $Bi_2Sr_2CaCu_2O_8$ (Ref. 9). Another interesting and related issue is the existence or nonexistence of an XRAY scattering peak at $Q = 2\pi/a_0(1/8,0)$ , because this is one of the indications that that CDW modulations are induced by $\lambda = 8a_0$ PDW order in the presence of DSC. Although such a peak has been observed for N($r$,E) both within vortex halos[42] and at zero-field[43] in $Bi_2Sr_2CaCu_2O_8$, attempts to detect it by XRAY measurements on $YBa_2Cu_3O_7$ at zero-field have not yet succeeded. Importantly, one of the revelations of our PDW+DSC model is that only a tiny $Q = 2\pi/a_0(1/8,0)$ charge modulation peak is predicted (black in Fig. 2b), whose intensity may be below present XRAY detection limits. Alternatively, the PDW in YBa$_2$Cu$_3$O$_7$ may be fluctuating and thus unobservable[44]. A distinct point is that a uniform DSC+PDW state should have clear spectroscopic signatures visible in angle-resolved photoemission (ARPES)[24,28]. And, to first order, the PDW+DSC model and ARPES correspond well, because the strong antinodal gap due to the PDW is empirically indistinguishable from the antinodal "pseudogap" reported by virtually all ARPES studies. But some fine features the PDW+DSC model predicts for $A(k,\omega)$ have not yet been seen by ARPES, presumably due to the short-range nature of the static PDW+DSC patches. Future theoretical modeling studies of disordered PDWs will be necessary to explore these issues. Lastly, one may wonder whether a 4a$_0$ periodic CDW state coexisting with a $d$-wave superconductor could reproduce the N($r$,E) and Z($r$,E) data or the of $\Lambda(q,\Delta_0)$ QPI signatures, as well as our PDW+DSC model. In that regard, we find that one must initialize the RMFT equations with a modulating pair-field to converge to a non-uniform state and that any other type of initialization, for example a 4a$_0$ periodic CDW, converges only to uniform DSC. More generally, Ginzburg-Landau models based on a large DSC order parameter coexisting with a CDW of wavevector $Q_{CDW}$ (and possibly an induced PDW at $Q_{CDW}$), do not have PDW at $Q_{CDW}/2$. Moreover, when we use our RMFT approach to study such CDW+DSC driven states, we find in two cases that the predicted $Z(r,E)$ spectra show very poor correlation with equivalent experiments (SI Section D). Therefore, the observation of 8a$_0$ periodic modulations in Z($r,E$) and $\Delta_p(r)$ along with a particle-hole symmetric kink at low energies, all of which are characteristics of the RMFT based PDW+DSC model but not of the others, favor the interpretation based on a pair density wave coexisting with superconductivity.

*Discussion and Conclusions*

12    To recapitulate, we have developed a strong-coupling mean-field theory for a coexisting $d$-wave superconductor and pair density wave ($\lambda = 8a_0$) and made detailed comparisons of its predictions with experimental SISTM data from Bi$_2$Sr$_2$CaCu$_2$O$_8$. To allow valid quantitative comparison to such experiments, the atomic-scale tunneling characteristics of the PDW+DSC state at the BiO termination layer of the crystal were predicted utilizing a recently developed Wannier-function based method. Then, from the $8a_0$ scale down to inside the $CuO_2$ unit cell, we find that the predictions (Fig. 2a-d, Fig. 3a-f) correspond strikingly well with the highly complex electronic structure patterns observed by SISTM (Fig. 3g-l) in the $p<p^*$ broken-symmetry state of $Bi_2Sr_2CaCu_2O_8$. Indeed, the PDW+DSC model explains simply the microscopic origins of many enigmatic characteristics of this broken-symmetry state, including the $\lambda = 4a_0$ $d$-symmetry form factor $N(r, E)$ modulations[6-10,12] and the concentration of their amplitude maxima[8,9,12,13] surrounding $|E| = \Delta_1$. Further, by considering scattering of $k$-space quasiparticles, we explore the QPI "fingerprint" $\Lambda(q, \Delta_0)$ of the PDW+DSC model (Fig. 5a), and find it too in strong, detailed agreement with measured $\Lambda(q, \Delta_0)$ (Fig. 5c). Hence, the antinodal pseudogap that dominates the experimental data $\Lambda(q, \Delta_0)$ (Fig. 5c) plausibly corresponds to the antinodal gap of the PDW in the model $\Lambda(q, \Delta_0)$ (Fig. 5a). Further, the transition at $p \approx 18\%$ in the model from the $\Lambda(q, \Delta_0)$ characteristic of a PDW+DSC state (Fig. 5a) to that of a pure d-wave superconductor (Fig. 5b) corresponds qualitatively with the experimentally observed[11] transition in $\Lambda(q, \Delta_0)$ at $p \approx p*$ (Fig. 5c,d ). Consequently, the disappearance at a critical hole-density $p^*$ of both the $r$-space symmetry-breaking structures (Fig. 3a-f) and the concomitant $q$-space QPI signatures (Fig. 5c) are all features that occur in the PDW+DSC model at the hole-density where the PDW disappears. Overall, the agreement between our PDW+DSC model and the plethora of experimental characteristics is consistent with a picture in which a disordered $\lambda = 8a_0$ PDW+DSC state exists for $p < p*$ in $Bi_2Sr_2CaCu_2O_8$, the $\lambda = 4a_0$ charge modulations observed by XRAY scattering[5,39] are a consequence of this state, the cuprate pseudogap coincides with the antinodal gap of the coexisting PDW, and the cuprate $p \approx p^*$ critical point is due to disappearance of the PDW.

*Acknowledgements:* The authors acknowledge and thank D. Agterberg, E.-A. Kim, M. Norman, S.A. Kivelson and Y. Wang for very helpful discussions and advice. P.C. and P.J.H. acknowledge support from NSF-DMR-1849751. P.C. acknowledges the research grant PDF/2017/002242 from SERB, DST, India. S.U. and H.E. acknowledge support from a Grant-in-aid for Scientific Research from the Ministry of Science and Education (Japan) and the Global Centers of Excellence Program for Japan Society for the Promotion of Science. S.H.J. and J.L. acknowledge support from the Institute for Basic Science in Korea (Grant No. IBS-R009-G2), the Institute of Applied Physics of Seoul National University, National Research Foundation of Korea (NRF) grant funded by the Korea government (MSIP) (No. 2017R1A2B3009576). S.H.J. also acknowledges support from the BK21 Plus Project. K.F. and Z.D. acknowledge support from the U.S. Department of Energy, Office of Basic Energy Sciences, under contract number DEAC02-98CH10886; J.C.S.D. acknowledges support from the Moore Foundation's EPiQS Initiative through Grant GBMF4544, from Science Foundation Ireland under Award SFI 17/RP/5445 and from the European Research Council (ERC) under Award DLV-788932.

*Author Contributions:* P.C., P.J.H and J.C.S.D. designed the project. S.H.J., Z.D., S.D.E., M.H.H., J.L. and K.F. carried out the experiments and data analysis. H.E. and S.U. synthesized samples. P.C. and P.J.H carried out theoretical analysis. J.C.S.D. and P.J.H supervised the investigation and wrote the paper with key contributions from P.C., S.H.J. and K.F. The manuscript reflects the contributions of all authors.

*Author Information* The authors declare no competing financial interests. Readers are welcome to comment on the online version of the paper. Correspondence and requests for materials should be addressed to P.J.H and J.C.S.D.

**FIGURE CAPTIONS**

**FIG. 1 Broken-symmetry Phase Coexisting with Cuprate Superconductivity**

a) Schematic phase diagram of lightly hole-doped $CuO_2$. The Mott insulator phase with long range antiferromagnetic order disappears quickly with increasing hole doping $p$ to be replaced by the pseudogap phase (PG). The d-wave superconductivity (DSC) coexists with PG phase below some critical hole density p*, and persists as a unique state at p>p*. Within the PG phase, the DW modulations have been reported.

b) Measured $Z(\boldsymbol{r}, E=\Delta_1)$ at the BiO termination layer of $Bi_2Sr_2CaCu_2O_8$ for p≈0.1 The locations of Cu atoms of the $CuO_2$ plane in the same field of view are shown on the topograph T(r) above it.

c) Measured differential conductance averaged over the same field of view as b. The black, dashed lines identify two characteristic energies $\Delta_1$ and $\Delta_0$.

**FIG. 2 Characteristic features of our model for a unidirectional PDW+DSC state**

(a) Variation of hole density (δ) and d-wave gap order parameter (Δ) with lattice sites i as obtained from the self-consistent solution of the extended t-J model for parameter set doping $x$=0.125,next nearest neighbor (NNN) hopping $t'$ = -0.3, exchange interaction $J$=0.3, and temperature T=0.04. Energy scale is presented in units of nearest-neighbor (NN) hopping $t$.

(b) Variation of hole density (δ) and d-wave gap order parameter (Δ) with wavevector $\boldsymbol{q}$ obtained by Fourier transforming corresponding lattice-space quantities shown in (a). Hole density and gap order parameter show largest modulating components at $|\boldsymbol{q}|$ = 0.25 and $|\boldsymbol{q}|$ =0.125, respectively. But the predicted $|\boldsymbol{q}|$ =0.125 component of charge density modulation δ(q) is extremely small, due to the small uniform component of the gap order parameter. Wavevectors are presented in units of $2\pi/a_0$, where $a_0$ is the lattice constant in the $CuO_2$ a-b plane.

(c) Density wave form-factors obtained from the hole-density and NN bond order parameter (χ). The d-symmetry form factor is dominant at all wavevectors, with predominant intensity at $|\boldsymbol{q}|$=0.25.

(d)     Continuum local density of states N(r,E) at Cu positions along Cu-$O_x$ direction over a period of PDW ($8a_0$). Here $\Delta_1 \approx 0.3$ and $\Delta_0 \approx 0.07$ correspond to the gap-scale associated with the PDW component and uniform DSC component of the PDW+DSC state, respectively.

**FIG 3. $Z(r,\omega)$ in PDW+DSC state compared to corresponding experimental data**

(a)-(f)  Predicted $Z(r,\omega)$ maps in PDW+DSC state for the terminal BiO layer, at energies $|\omega|=0.4\Delta_1, 0.6\Delta_1, 0.8\Delta_1, \Delta_1, 1.2\Delta_1$, and $1.4\Delta_1$, respectively.

(g)-(i)  Experimentally measured $Z (r,E/\Delta_1)$ maps at the terminal BiO layer, for bias voltages corresponding to energies in (a)-(f), respectively.

(m)     Cross-correlation coefficient between theoretical $Z(r,\omega)$ images and experimental $Z(r,E/\Delta_1)$ map images as a function of energy/voltage, showing very strong correspondence between them in the region ($0.5\Delta_1$ -$1.5\Delta_1$, marked with vertical black dashed lines) around PDW energy gap scale. Direct examination of a typical pair, for example (e) and (k), show why the cross-correlation coefficient is so high. A wide variety of minute details, including the distinct broken rotational symmetry inside each specific $CuO_2$ unit cell, the d-symmetry modulations of this broken symmetry over eight unit cells, and the bond-centered register of this $8a_0$ unit to the $CuO_2$ lattice, correspond strikingly between experiment and theory.

**FIG. 4 Gapmap $\Delta_p(r)$ derived from coherence-peak energy in PDW+DSC state**

(a)     Gapmap in PDW+DSC state obtained by identifying coherence peak energy $\Delta_p(r)$ for $\omega>0$ at all intra-unit cell points over an area of 8x12 unit cells. Colorbar is given in the units of 't'.

(b)     Gapmap $\Delta_p(r)$ obtained by following the same procedure as in (a), but for experimentally measured g(r,V) spectra of a representative domain as shown.

(c)     Gap $\Delta_p(r)$ averaged along y-direction (black) obtained in the PDW+DSC model.

(d)     Gap $\Delta_p(r)$ averaged along y-direction (black) obtained from the g(r,V) spectra.

**FIG 5. Comparison** of $\Lambda(q, \Delta_0)$ **predictions/data in PDW+DSC and pure DSC states**

(a)     Predicted $\Lambda(q, \Delta_0)$ map in PDW+DSC state at doping x=0.125.

(b)     Predicted $\Lambda(q, \Delta_0)$ map in a uniform d-wave superconductor state at doping x=0.23.

(c)     Experimental $\Lambda(q, \Delta_0)$ map for underdoped $Bi_2Sr_2CaCu_2O_8$ at p=0.08$\pm$0.1. In principle $T(q, eV = \Delta_0)$, the Fourier transform of a topographic image $T(r, eV = \Delta_0)$, would be logarithmically sensitive to the same information as $\Lambda(q, \Delta_0)$ . But because it also contains such intense signals from a variety of other phenomena, is has proven difficult to use that approach for QPI studies.

(d)     Experimental $\Lambda(q, \Delta_0)$ map for overdoped $Bi_2Sr_2CaCu_2O_8$ at p=0.23.

**Figure 1**

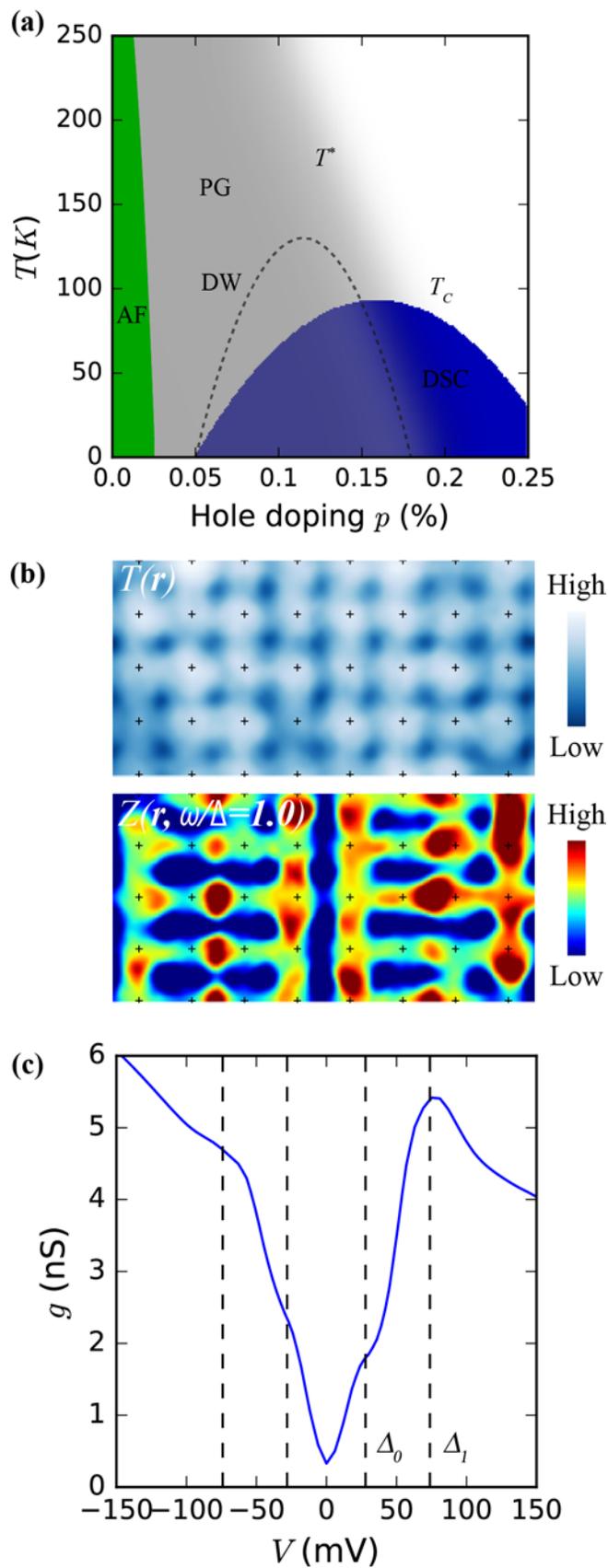

**Figure 2**

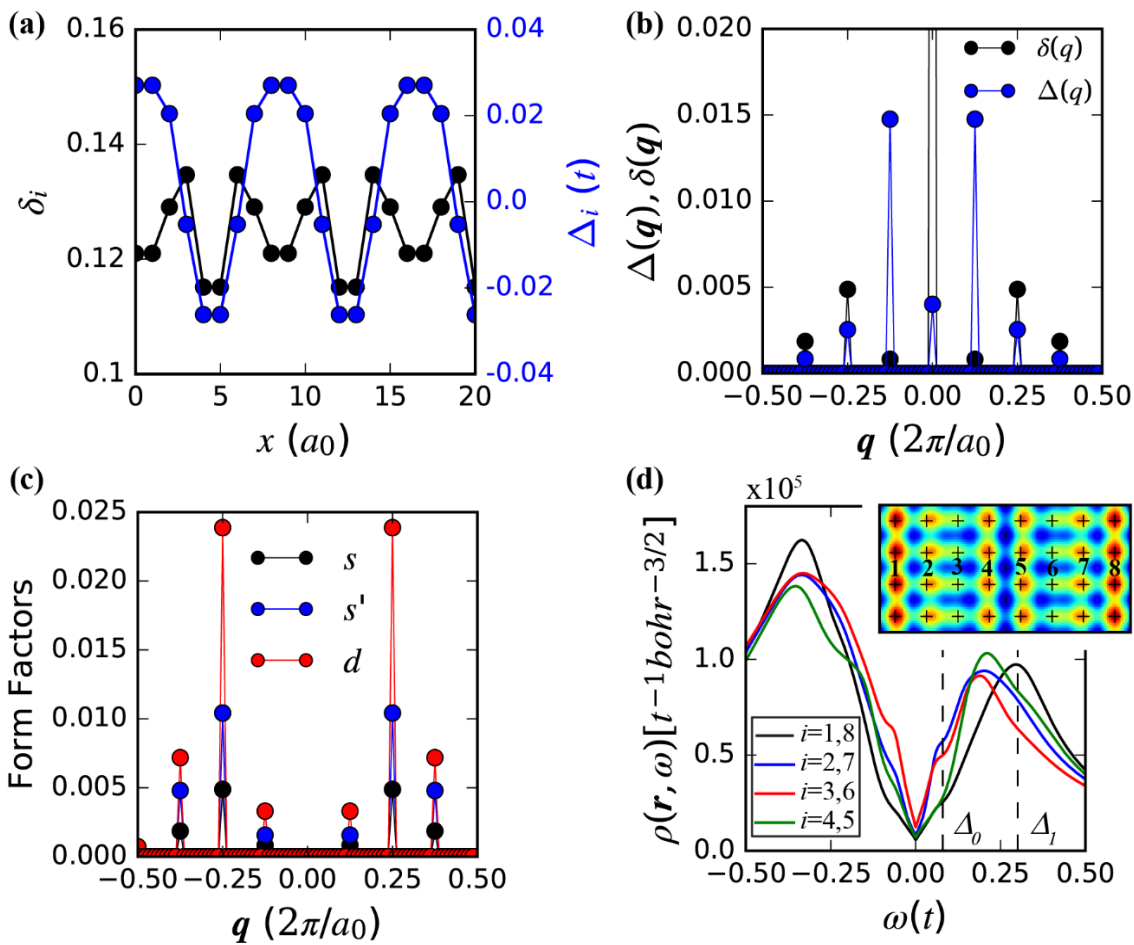

**Figure 3**

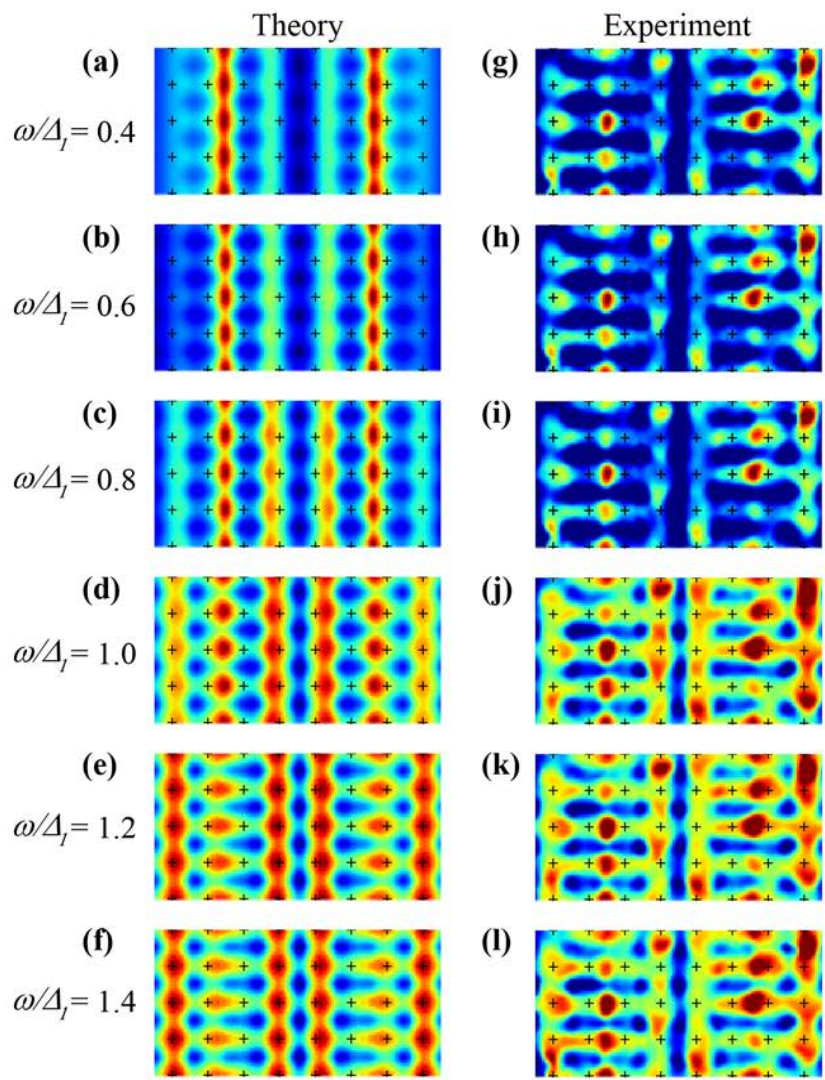

Theory      Experiment

(a)    $\omega/\Delta_1 = 0.4$    (g)

(b)    $\omega/\Delta_1 = 0.6$    (h)

(c)    $\omega/\Delta_1 = 0.8$    (i)

(d)    $\omega/\Delta_1 = 1.0$    (j)

(e)    $\omega/\Delta_1 = 1.2$    (k)

(f)    $\omega/\Delta_1 = 1.4$    (l)

Low ▬ High

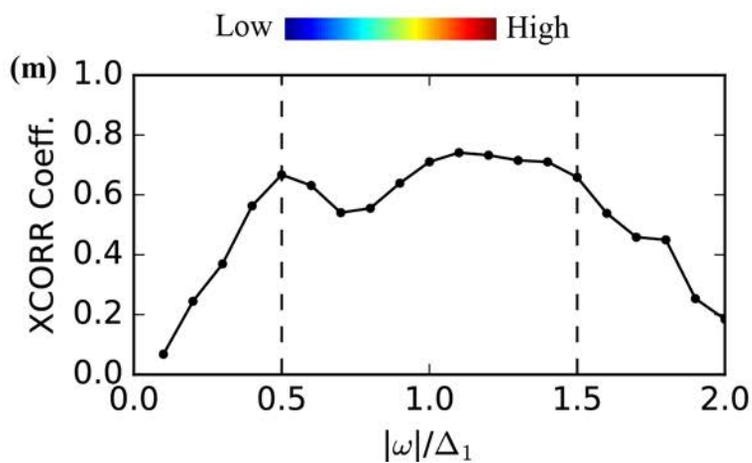

(m)

# Figure 4

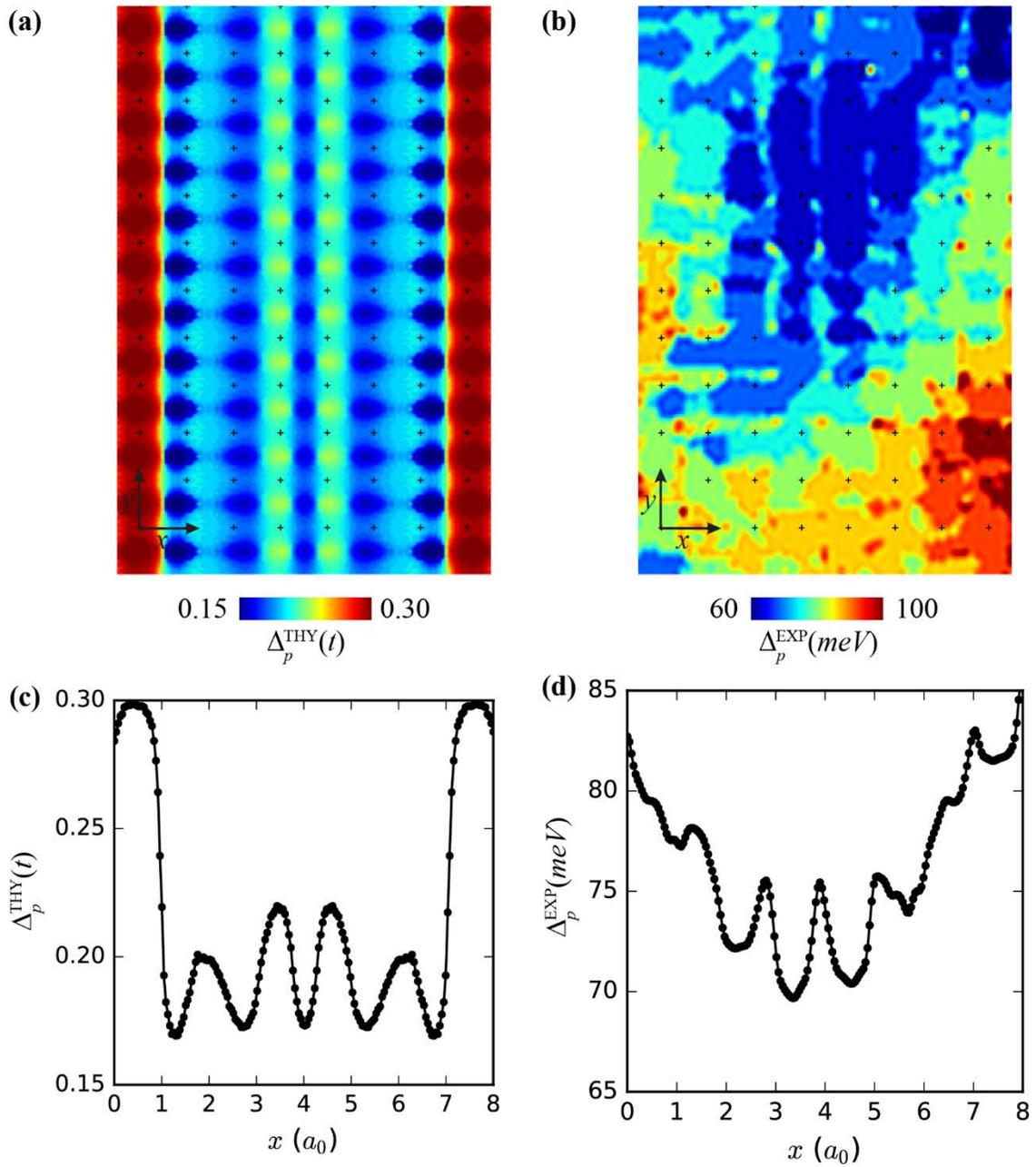

**Figure 5**

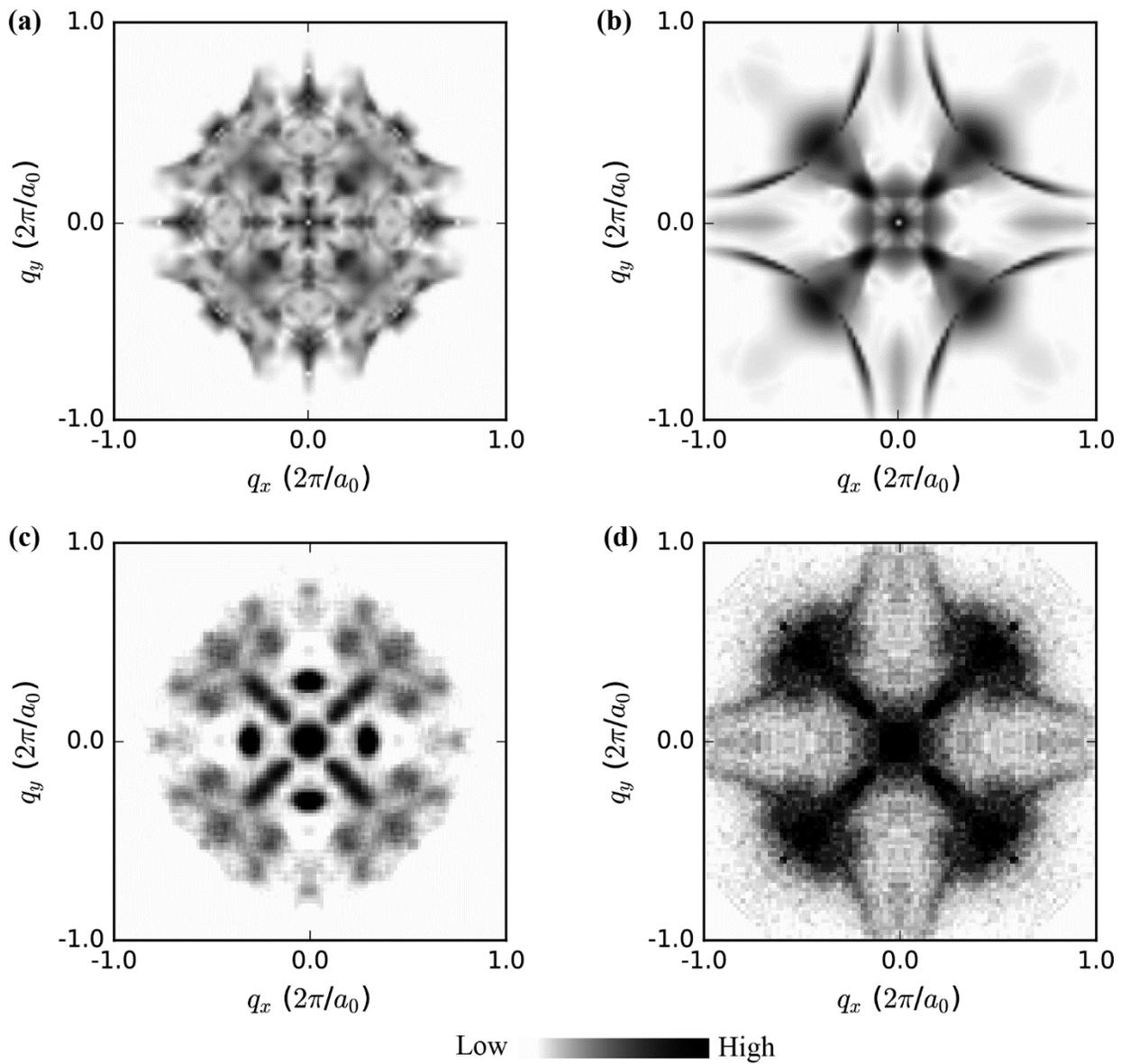

**Supplementary Information**

# Atomic-scale Electronic Structure of the Cuprate Pair Density Wave State Coexisting with Superconductivity

Peayush Choubey, Sang Hyun Joo, K. Fujita, Zengyi Du, S. D. Edkins, M.H. Hamidian,

H Eisaki, S. Uchida, A. P. Mackenzie, Jinho Lee, J.C. Séamus Davis and P.J. Hirschfeld

## Section A Renormalized mean-field theory of the extended t-J model

For CuO$_2$ plane the *t-J* model is given by

$$H = -\sum_{(i,j),\sigma} P_G t_{ij}(c_{i\sigma}^\dagger c_{j\sigma} + h.c.) P_G + J\sum_{<i,j>} \mathbf{S}_i \cdot \mathbf{S}_j, \tag{S1}$$

where the hopping matrix elements $t_{ij}$ are taken to be $t$ for Cu-nearest neighbors (NN) and $t'$ for next nearest neighbors (NNN), and $J$ is their Heisenberg exchange. The operator $P_G$ projects out doubly occupied sites from the Hilbert space. The Gutzwiller approximation to this model of strongly correlated electrons replaces the exact projection $P_G$ with classical counting factors $g^t$ and $g^s$ depending on the allowed number of configurations of the degrees of freedom on a given site. This is determined by comparing expectation values of the kinetic and potential energies with their values taken in the Gutzwiller variational wavefunction.

This approach can be generalized to a include set of *local* renormalization factors, such that the renormalized Hamiltonian now reads,

$$H = -\sum_{(i,j),\sigma} g_{ij}^t t_{ij}(c_{i\sigma}^\dagger c_{j\sigma} + h.c.) + J\sum_{<i,j>} \left[ g_{ij}^{s,z} S_i^z S_j^z + g_{ij}^{s,xy}\left(\frac{S_i^+ S_j^- + S_i^- S_j^+}{2}\right)\right] \tag{S2}$$

where $<i,j>$ and $(i,j)$ represent only NN, and both NN and NNN sites, respectively. Gutzwiller factors $g^t$ and $g^s$ then depend on the local mean fields. Once calculated in a self-consisted procedure, all physical observables can be obtained, and the locality of the fields allows the consideration of spatially modulated solutions to the mean field equations.

Mean-field decoupling of the renormalized Hamiltonian in Eq. S2 in pairing and density channels, and subsequent minimization of the ground state energy $E_g = <\Psi_0|H|\Psi_0>$ with respect to the ground state wavefunction $|\Psi_0>$ leads to the following mean-field Hamiltonian [1]

$$H = -\sum_{(i,j),\sigma} \epsilon_{ij} c_{i\sigma}^\dagger c_{j\sigma} + h.c. + \sum_{<i,j>,\sigma} \sigma D_{ij}^* c_{i\sigma} c_{j\bar\sigma} + h.c. - \sum_{i,\sigma} \mu_{i\sigma} n_{i\sigma}, \tag{S3}$$

where,

$$\epsilon_{ij\sigma} = -g_{ij}^t t_{ij} - \delta_{ij,<ij>} \frac{3}{4} J g_{ij}^s \chi_{ij\sigma}^* \tag{S4}$$

$$D_{ij\sigma} = -\delta_{ij,<ij>} \frac{3}{4} J g_{ij}^s \Delta_{ij\sigma} \tag{S5}$$

$$\mu_{i\sigma} = \mu + \frac{3}{4} J \sum_{j\sigma'} \left( |\Delta_{ij\sigma'}|^2 + |\chi_{ij\sigma'}|^2 \right) \frac{dg_{ij}^s}{dn_{i\sigma}} + t_{ij} \sum_{j\sigma'} \left( \chi_{ij\sigma'} + \chi_{ij\sigma'}^* \right) \frac{dg_{ij}^t}{dn_{i\sigma}} \tag{S6}$$

Here, $\delta_{ij,<ij>} = 1$ for NN sites and 0 otherwise. Various mean-fields appearing in the above equation are defined as

$$\Delta_{ij\sigma} = \sigma < \Psi_0 | c_{i\sigma} c_{j\bar{\sigma}} | \Psi_0 > \tag{S7}$$

$$\chi_{ij\sigma} = < \Psi_0 | c_{i\sigma}^\dagger c_{j\sigma} | \Psi_0 > \tag{S8}$$

$$m_i = < \Psi_0 | S_i^z | \Psi_0 > \tag{S9}$$

$$\delta_i = 1 - < \Psi_0 | \sum_\sigma n_{i\sigma} | \Psi_0 > \tag{S10}$$

In writing Eq. S1, we have considered only paramagnetic states ($m_i = 0$) as we are interested in charge ordering without spin ordering. This significantly simplifies the expressions for Gutzwiller factors [1] as given below

$$g_{ij\sigma}^t = g_{ij}^t = g_i^t g_j^t; \ g_i^t = \sqrt{\frac{2\delta_i}{1-\delta_i}} \tag{S11}$$

$$g_{ij}^{s,z} = g_{ij}^{s,xy} = g_{ij}^s = g_i^s g_j^s; \ \ g_i^s = \frac{2}{1+\delta_i} \tag{S12}$$

Applying a Bogoliubov transformation diagonalizes the mean-field Hamiltonian in Eq. S3, and leads to the following BdG equation

$$\sum_j \begin{pmatrix} \epsilon_{ij\uparrow} & D_{ij\uparrow} \\ D_{ji\uparrow}^* & -\epsilon_{ij\downarrow} \end{pmatrix} \begin{pmatrix} u_j^n \\ v_j^n \end{pmatrix} = E_n \begin{pmatrix} u_i^n \\ v_i^n \end{pmatrix} \tag{S13}$$

which has to be solved self-consistently as the matrix elements depend on the mean-fields, which, in turn, are determined from the eigenvalues ($u_i^n, v_i^n$) and eigenvectors $E_n$. To obtain bond-centered PDW+dSC state as a self-consistent solution of Eq. S13, we initialize it with modulating bond-centered pair-field, keeping hole density and bond-field uniform, as follows

$$\Delta_{i,i+\hat{x}} = \Delta_0 + \Delta_Q \cos(\boldsymbol{Q_P} \cdot \boldsymbol{R_i}) \tag{S14}$$

$$\Delta_{i,i+\hat{y}} = -\Delta_0 - \Delta_Q \cos\left[\boldsymbol{Q_P} \cdot \left(\boldsymbol{R_i} - \frac{\hat{x}}{2}\right)\right] \tag{S15}$$

where $\boldsymbol{Q_P} = (1/8,0)2\pi/a_0$.

Fig. S1(a) shows the doping dependence of the energy per site ($E_t$) in the DSC and PDW+DSC state at T=0. PDW+DSC is found to be a stable solution above doping $x \approx 0.04$ and below $x = x_c \approx 0.18$, whereas the dSC states exists for a much wider doping range as seen in the inset. At very low doping ($x < 0.04$), it's very difficult to obtain a converged PDW+DSC solution, presumably due to large changes in $\frac{dg_{ij}^t}{dn_{i\sigma}}$ caused by small deviations in hole density[5]. For dopings $x > x_c$, the self-consistent solution converges to the DSC state, even if initialized with PDW+DSC configuration. At $x = x_c$, the change in PDW order parameter ($\Delta_Q$) occurs via a first-order transition, see Fig. S1(b). The dSC state has the lowest energy per site at all dopings considered, however the difference between PDW+DSC and DSC is in the range $[0.0009t, 0.0081t]$ (or $[0.36, 3.24]meV$ taking $t = 400\ meV$), which is so tiny (O(1meV)) [1], that it may be stabilized by a variety of means, including disorder. In the main text, we have assumed it to be true and explored the consequences for various STM observables.

The PDW order parameter obtained here has dominant $d$-form factor. However, there is a small admixture of $s'$-form factor since the PDW wavevector $\boldsymbol{Q}_P$ breaks the four-fold symmetry of the lattice. The bond-field $\chi_{ij\sigma}$ inherits the form-factor structure from PDW via self-consistency. In Fig. 2(c), we show the density wave form-factors obtained using the self-consistent charge and bond-fields as follows [2, 3]. We first obtain the NN bond order parameters $\bar{\chi}_{i,i+\hat{x}(\hat{y})}$, and hole density $\delta_i$, and, subsequently, obtain the s-, s'-, and d-form factors in reciprocal space as below:

$$\bar{\chi}_{i,i+\hat{x}(\hat{y})} = \frac{1}{2}\sum_\sigma g_{i,i+\hat{x}(\hat{y}),\sigma}^t \chi_{i,i+\hat{x}(\hat{y}),\sigma} + \text{H. c.} \tag{S16}$$

$$d(\boldsymbol{q}) = FT(\bar{\chi}_{i,i+\hat{x}} - \bar{\chi}_{i,i+\hat{y}})/2 \tag{S17}$$

$$s'(\boldsymbol{q}) = FT(\bar{\chi}_{i,i+\hat{x}} + \bar{\chi}_{i,i+\hat{y}})/2 \tag{S18}$$

$$s(q) = FT(1 - \delta_i) \tag{S19}$$

Here, *FT* represents the Fourier transform over lattice space. In Fig. 2(c), we show the magnitude of the $\boldsymbol{q} \neq \boldsymbol{0}$ components of the form-factors.

**Section B Wannier transformation from CuO₂ plane electronic structure to N(r,E)**

Using the eigenvalues and eigenvectors of the BdG matrix (Eq. S13), lattice Green's functions can be calculated using the following formula

$$G_{ij}(\omega) = g_{ij}^t \sum_n \frac{u_i^n u_j^{n*}}{\omega - E_n + i0^+} \tag{S20}$$

where, $0^+$ is a small artificial broadening chosen to be 0.005t, and the sum runs over all the eigenvalues. The total LDOS at a lattice site is simply given by the imaginary part of the diagonal Green's function:

$$N_i(\omega) = -\frac{2}{\pi} Im[G_{ii}(\omega)], \tag{S21}$$

where the factor 2 accounts for spin degeneracy.

Fig. S2(a) shows the lattice LDOS thus obtained in the PDW+DSC state for the same parameter set as in Fig. 2. The LDOS displays a number of features (including peaks at $\Delta_0$ and $\Delta_1$, among others). Sharp features present at higher energies will be smeared by inelastic scattering, which has been shown to be necessary to account for the shape of spectra in underdoped cuprates [6]. We incorporate the effects of linear inelastic scattering by replacing the constant artificial broadening $i0^+$ by a bias-dependent term $i0^+ + i\Gamma$, where $\Gamma = \alpha|\omega|$. Using the experimental fits presented in Ref. [6], we set $\alpha = 0.25$. Fig. S2(b) shows the lattice LDOS incorporating the linear inelastic scattering. All LDOS (and quantities directly related to it) results presented in the main text have been obtained after accounting for the inelastic scattering.

For a proper theoretical interpretation of the experimentally measured $N(\boldsymbol{r}, E)$, one must calculate LDOS at the STM tip position $\boldsymbol{r}$. This can be achieved by a basis transformation from lattice to continuum space where Wannier functions centered at the lattice sites serve as matrix elements of the transformation [4]. The resulting Green's function and LDOS in the continuum basis is given as

$$G(\boldsymbol{r}, \boldsymbol{r}'; \omega) = \sum_{ij} W_i(\boldsymbol{r}) G_{ij}(\omega) W_j^*(\boldsymbol{r}') \tag{S22}$$

$$N(\boldsymbol{r}, \omega) = -\frac{2}{\pi} Im[G(\boldsymbol{r}, \boldsymbol{r}; \omega)] \tag{S23}$$

Here, $W_i(\boldsymbol{r})$ is the Wannier function centered at site $i$. To compute the continuum LDOS at a typical STM-tip height, we have employed Cu-$3d_{x2-y2}$ Wannier function obtained using Wannier90 package, same as in Ref. [3, 7] and very similar to that in Ref. [8].

Using continuum LDOS at Ox and Oy sites, the density wave form factors can be obtained as a function of bias. Similar to the experimental procedure[9], we first obtain the sublattice Z-maps $Cu(\boldsymbol{r}, E)$ containing only $Z(\boldsymbol{r}, E)(= \frac{N(\boldsymbol{r}, E)}{N(\boldsymbol{r}, -E)})$ at copper sites and $O_x(\boldsymbol{r}, E)$ and $O_y(\boldsymbol{r}, E)$, containing only $Z(\boldsymbol{r}, E)$ at the $x/y$-axis oxygen sites. Then we Fourier transform the sublattice Z-maps to obtain $Cu(\boldsymbol{q}, E)$, $O_x(\boldsymbol{q}, E)$ and $O_y(\boldsymbol{q}, E)$, and compute the form factors using the following equations.

$$d(\boldsymbol{q}, E) = |(O_x(\boldsymbol{q}, E) - O_y(\boldsymbol{q}, E))|/2 \tag{S24}$$

$$s'(\boldsymbol{q}, E) = |(O_x(\boldsymbol{q}, E) + O_y(\boldsymbol{q}, E))|/2 \qquad (S25)$$

$$s(\boldsymbol{q}, E) = Cu(\boldsymbol{q}, E) \qquad (S26)$$

Fig. S3 shows the bias-dependence of the form factors in PDW+DSC state for $\boldsymbol{q} = 2\boldsymbol{Q_P} = (1/4,0)2\pi/a_0$. The d-form factor remains dominant at higher energies, as seen in the experiment[9]. The largest value of d-form factor relative to s'-form factor occurs near $E \approx \Delta_1$. Moreover, as the d-form factor begins to dominate, the phase difference between $O_x$ and $O_y$ modulation approaches π. A pure d-form factor density wave, by definition, implies that modulations at $O_x$ and $O_y$ sites are out of phase. In our case, the admixture of s'-form factor leads to small deviations from π.

**Section C Bogoliubov quasiparticle scattering interference**

To compute Bogoliubov quasiparticle scattering interference (BQPI), we solve single-impurity scattering problem in the extended *t-J* model self-consistently. In presence of a point-like potential scatterer located at the lattice site *i\**, the following Hamiltonian describes the system.

$$H = H_{t-J} + H_{imp},$$

$$H_{imp} = V_{imp} \sum_{\sigma} n_{i^*\sigma}, \qquad (S27)$$

where, $H_{t-J}$ is given by Eq. S1, and $V_{imp}$ is the impurity potential. Inclusion of this impurity Hamiltonian modifies the mean-field equation S6 as $\mu_{i\sigma} \rightarrow \mu_{i\sigma} - V_{imp}\delta_{ii^*}$, (here $\delta_{ij}$ is the Kronecker delta function) leaving everything else unchanged. The modified BdG equations are then solved self-consistently to yield DSC and PDW+DSC states in presence of an impurity at the origin. Subsequently, we compute continuum LDOS $N(\boldsymbol{r}, E)$ using the procedure described in Section B. To compute BQPI, we obtain Fourier transform of the $Z(\boldsymbol{r}, E)$ map at each $E$ to obtain $Z(\boldsymbol{q}, E)$. Since, the continuum LDOS has intra-unit cell precision, BQPI at wavevectors in higher Brillouin zones are available, in contrast to the lattice LDOS-based calculations, which provides information about the wavevectors in the first Brillouin zone only. Fig. S4(a) shows as obtained $Z(\boldsymbol{q}, E)$ map at a low bias $E = 0.1\Delta_1 (= 0.03t)$ in the PDW+DSC state. The largest intensity occurs at the non-dispersing PDW Bragg wavevectors $\boldsymbol{q} = n\boldsymbol{Q_P}, n = \pm 0, \pm 1, \dots, \pm 7$, making it hard to see the other dispersing wavevectors. We assume that the discommensuration present in the real material [10] will significantly reduce the intensity at the Bragg wavevectors, and suppress it "by hand" in the theoretical BQPI maps, as shown in Fig. S4(b). Moreover, to simulate the presence of domains with unidirectional density modulations along both x- and y-axes, we symmetrize the BQPI maps by adding their 90°-rotated versions, as shown in Fig. S4(c). All $\Lambda$-maps in PDW+DSC state shown in the main-text (Fig. 4(a)) and hereafter are obtained by following this procedure of Bragg peak suppression and four-fold symmetrization.

The integrated $Z(\boldsymbol{q}, E)$ map, i.e. the $\Lambda(\boldsymbol{q})$ map provides an efficient way to identify the characteristic features of Cooper pairing in both underdoped and overdoped Cuprates [11]. The overdoped BSCCO shows a full Fermi surface-tracing trajectory, which is a hallmark of simple d-wave pairing and corresponding octet model [11]. In the case of underdoped BSCCO, only a part of the Fermi surface (an "arc") is observed when the $Z(\boldsymbol{q}, E)$ maps are integrated up to the uniform d-wave pairing gap scale $\Delta_0$. No additional feature emerges if the upper limit of sum is set beyond $\Delta_0$. We verify this by defining the $\Lambda$-map as a function of upper energy cut-off $\Delta_c$:

$$\Lambda(\boldsymbol{q}, \Delta_c) = \sum_{E \cong 0}^{\Delta_c} Z(\boldsymbol{q}, E) \qquad \text{(S28)}$$

In Fig. S5, we plot $\Lambda$-maps at various cut-offs. The cut-off for the left-most map $\Delta_c = 0.2\Delta_1 (= 0.06t)$ is very close to $\Delta_0 (= 0.07t)$ and hence it is quite similar to the map presented in Fig. 4(a) in the main-text. As we increase the cut-off, no new feature emerges. In particular, the length of the arc like feature remains almost same throughout, demonstrating that the characteristics of the integrated BQPI maps is determined mostly by the low-energy scale $\Delta_0$.

In Figure S6, we present the doping dependence of the BQPI in the pure $d$-wave state, with Fermi surface and gap calculated self-consistently within the RMFT. It is interesting to note that Figure 5(b) in the main text, the theoretical prediction for a d-wave gap $\Delta_{\boldsymbol{k}} \sim \cos k_x - \cos k_y$ using a phenomenological Fermi surface fit [8] to ARPES data with $x = 0.23$, compares much more favorably with experimental data at overdoping (Fig. 5d) than Fig. S6(b) as shown here. This suggests that the RMFT overestimates the persistence of correlations to the overdoped side of the phase diagram.

**Section D Significance of PDW order**

Very often, the STM and X-ray results on the charge ordered state in underdoped cuprates are interpreted in terms of uniform d-wave superconductivity coexisting with d-form factor charge density wave (CDW) with wavevector $(Q_c, 0) / (0, Q_c)$, where $Q_c \approx 0.25$ [12]. It is natural to ask whether the detailed STM results shown in the main text can have an alternative explanation in terms of such a DSC+CDW state. Unfortunately, it is not possible to answer this question within the same self-consistent framework since CDW+DSC state does not occur as a stable solution of the RMFT equations. However, non-self consistent model calculations, where bond- and pair-fields are tuned "by hand", can be utilized to compute various STM observables for comparison with the STM data. In particular, we have investigated the following two scenarios:

1. **Suppressing PDW part of PDW+DSC state:** We start with PDW+DSC self-consistent mean fields and replace the pair-field by that obtained for the uniform DSC state, keeping other mean-fields the same. The state thus created "by hand" is CDW+BDW+DSC. Here, CDW implies modulation of on-site potential ($\mu_{i\sigma}$) and BDW (bond density wave) implies modulation of NN bond mean-field ($\chi_{i,i+\hat{x}(\hat{y})}$) with a dominant d-form factor. This state exhibits the largest LDOS modulations at wavevector ($2Q_p$, 0) and smaller modulations at ($Q_p$, 0). It shows an absence of any low-energy kink in the spectrum and poor cross-correlation of Z-maps with the corresponding experimental data, as shown in Fig. S7 and S8, respectively.

2. **BDW+DSC state model:** Starting with uniform DSC state mean-fields, we add a bond density wave (BDW) in the form of d-form factor NN bond-fields $\chi_{i,i+\hat{x}(\hat{y})}$ "by hand":

$$\chi_{i,i+\hat{x}} = \Delta_{BDW} \cos(\boldsymbol{Q}_c \cdot \mathbf{R}_i) \tag{S29}$$

$$\chi_{i,i+\hat{y}} = -\Delta_{BDW} \cos\left[\boldsymbol{Q}_c \cdot \left(\mathbf{R}_i - \frac{\hat{x}}{2}\right)\right] \tag{S30}$$

Here, $\Delta_{BDW}$ and $\boldsymbol{Q}_c = (1/4, 0)2\pi/a_0$ are the BDW amplitude and wavevector, respectively. This model BDW+DSC state has a periodicity of 4 unit-cells and is often thought to be the most probable charge density wave candidate to explain the STM and X-ray data [12]. Tuning the BDW amplitude relative to DSC amplitude ($\Delta_{DSC}$), we find that the low-energy kink-like feature appears in the spectrum only when the two amplitudes are comparable, as shown in Fig. S9. Even in this parameter regime, the cross-correlation of Z-maps with the corresponding experimental data turns out to be poor, see Fig. S10.

In spite of the fact that aforementioned scenarios are non-self consistent and certainly not exhaustive, they strengthen the conclusion that a PDW driven charge ordering explains the detailed STM observables much better than the CDW driven mechanism. When seen together with the discovery of $\boldsymbol{Q}_p = (1/8, 0)2\pi/a_0$ peak in zero-field and in-field STM experiments [13], PDW+DSC scenario becomes much more probable.

**Section E  PDW gap map analysis from multiple samples/domains**

We investigated three different underdoped $Bi_2Sr_2CaCu_2O_{8+x}$ samples with $T_C$ of 20 K, 45 K and 50 K. Fig. S11 (a1-a5) show experimentally measured $Z(\boldsymbol{r}, E/\Delta_1 = 1)$ maps at the terminal BiO layer of PDW+DSC state. Gap maps are obtained by identifying the coherence peak energy, $\Delta_P(\boldsymbol{r})$, for $\omega > 0$ from the measured g($\boldsymbol{r}$,E) spectra in the domains and shown in Fig. S11 (b1-b5). Each gap maps is normalized by its spatial average, then averaged along $y$-direction to show its modulation along $x$-direction. Fig. S11 (c1-c5) display the $\Delta_P(\boldsymbol{r})$ modulation along $x$-direction of the domains. They clearly demonstrate that, although measured at quite different doping levels, an $8a_0$ periodic gap modulation in the measured gap maps exists robustly, and also maintains the same spatial phase in every domain. The average of the modulations along $x$-direction, $\Delta_{\mathrm{AVG}}(x) = \sum_{i=1}^{5} \Delta_{p,i}(x)/5$ over the five representative domains is shown in Fig. S11 (d), and as is the magnitude of the standard deviation (blue error bar).

## Section F  PDW+DSC predictions for scanned Josephson  tunneling microscopy

Using the $\mathrm{Bi_2Sr_2CaCu_2O_{8+\delta}}$ nanoflake-tip technology, on $\mathrm{Bi_2Sr_2CaCu_2O_{8+\delta}}$ samples similar to those studied here, the magnitude of the Josephson current $|I_J(\boldsymbol{r})|$ is found to modulate with wavelength $\approx 4\mathrm{a_0}$. In scanned Josephson tunneling microscopy (SJTM), the circuitry measures the magnitude of Josephson critical current magnitude  $|I_J|$ and thereby the rate of Cooper pair tunneling from both the homogenous DSC condensate and the PDW. Under the circumstances where the DSC order parameter exceeds that of the PDW, the $|I_J(\boldsymbol{r})|$ is not expected to pass through zero, nor does it in the extant experiments at $p \approx 0.17$. However, the PDW+DSC model presented here indicates that at lower $p$, when the PDW order parameter is the stronger of the two, then the $|I_J(\boldsymbol{r})|$ should modulate sufficiently to pass through zero with periodicity $4\mathrm{a_0}$, and this prediction will become the focus for future SJTM studies.

# SI Figure Captions

**Figure S1**

(a) Energy per site ($E_t$) at various hole-dopings ($x$) in DSC and PDW+DSC states at T=0. Inset shows the dome-shaped variation of DSC order parameter, with two vertical red lines showing the doping range where PDW+DSC state is stable. (b) Variation of d-wave gap order parameter ($\Delta$) with wavevector q just below (red) and above (blue) the critical doping $x_c$.

**Figure S2**

Lattice LDOS in PDW+DSC state over a period of PDW ($8a_0$), calculated using Eq. S21, for the same parameter set and numbering of lattice sites as in Fig. 2 of the main text (a) without and (b) with $\Gamma=\alpha|\omega|$ inelastic scattering ($\alpha=0.25$)[6].

**Figure S3**

Bias dependence of (a) form factors and (b) phase difference ($\Delta\varphi$) between modulations at Ox and Oy sites in the PDW+DSC state, obtained using Eq. S24-S26.

**Figure S4**

(a) BQPI $\boldsymbol{Z}(\boldsymbol{q},\boldsymbol{E})$-map at energy $\boldsymbol{E}=\boldsymbol{0}.\boldsymbol{1}\boldsymbol{\Delta_1}$(=0.03t) in PDW+DSC state obtained using parameters same as in Fig. 4 of the main-text. Largest intensity occurs at non-dispersing Bragg peaks $\boldsymbol{q}=\boldsymbol{n}\boldsymbol{Q_P},\boldsymbol{n}=\boldsymbol{0},\boldsymbol{1},\dots,\boldsymbol{7}$. (b) Same as in (a) with Bragg-peaks suppressed for a better visualization of dispersing wavevectors. (c) Symmetrized map obtained by taking the average of the map in (b) and its 90° rotated version.

**Figure S5**

$\Lambda$-Maps obtained at various upper energy cut-offs $\Delta_c$ (see Eq. S28) shown at the top of each panel in the units of PDW gap scale $\Delta_1$.

**Figure S6**

$\Lambda$-Maps in DSC state for t'=-0.3t (a)-(d) and t'=-0.6t (e)-(h) at various hole doping levels x shown on the top-left corner of each panel.

**Figure S7**

Continuum local density of states $\rho(r_i, E)$ at Cu positions $r_i$ along Cu-$O_x$ direction over a period of CDW ($8a_0$) in CDW+BDW+DSC state obtained by replacing the PDW+DSC state pair-field with uniform DSC state pair-field "by hand", keeping other mean-fields unchanged.

**Figure S8**

$Z(r,\omega)$ in CDW+BDW+DSC state compared to corresponding experimental data. (a)-(f) Predicted $Z(\boldsymbol{r},\omega)$ maps in CDW+BDW+DSC state for the terminal BiO layer, at energies $|\omega|=0.4\Delta_1$, $0.6\Delta_1$, $0.8\Delta_1$, $\Delta_1$, $1.2\Delta_1$, and $1.4\Delta_1$, respectively. (g)-(l) Experimentally measured $Z(\boldsymbol{r}, E/\Delta_1)$ maps at the terminal BiO layer, for bias voltages corresponding to energies in (a)-(f), respectively. (m) Cross-correlation coefficient between theoretical $Z(r,\omega)$ images and experimental $Z(\boldsymbol{r}, E/\Delta_1)$ map images as a function of energy/voltage, showing a poor correspondence between them.

**Figure S9**

Lattice LDOS spectrum over a period of CDW ($4a_0$) in BDW+DSC state, obtained by adding d-form factor bond field (see Eq. S29-S30) to uniform DSC state mean-fields, for various ratios of BDW and DSC amplitudes $\Delta_{BDW}/\Delta_{DSC}$.

**Figure S10**

$Z(r,\omega)$ in BDW+DSC state with $\Delta_{BDW}/\Delta_{DSC} = 1$ compared to corresponding experimental data. (a)-(f) Predicted $Z(\boldsymbol{r},\omega)$ maps in BDW+DSC state for the terminal BiO layer, at energies $|\omega|=0.4\Delta_1$, $0.6\Delta_1$, $0.8\Delta_1$, $\Delta_1$, $1.2\Delta_1$, and $1.4\Delta_1$, respectively. (g)-(l) Experimentally measured $Z(\boldsymbol{r}, E/\Delta_1)$ maps at the terminal BiO layer, for bias voltages corresponding to energies in (a)-(f), respectively. (m) Cross-correlation coefficient between theoretical $Z(\boldsymbol{r},\omega)$ images and experimental $Z(\boldsymbol{r}, E/\Delta_1)$ map images as a function of energy/voltage, showing a poor correspondence between them.

**Figure S11**

(a1-a5) The measured $Z(\boldsymbol{r}, E/\Delta_1 = 1)$ maps at the terminal BiO layer of PDW+DSC state in three different samples: UD20 ($T_c$=20K), UD45 ($T_c$=45K) and UD50 ($T_c$=50K). (b1-b5) Gapmaps, $\Delta_P(\boldsymbol{r})$, of the domains in (a1-a5). (c1-c5) Gap modulation along x-direction of each domain. Each gap map is normalized by its spatial average and averaged along y-axis. (d) Average of plots in (c1-c5). Blue error bars represent the magnitude of standard deviation of (c1-c5).

**Figure S1**

**(a)**

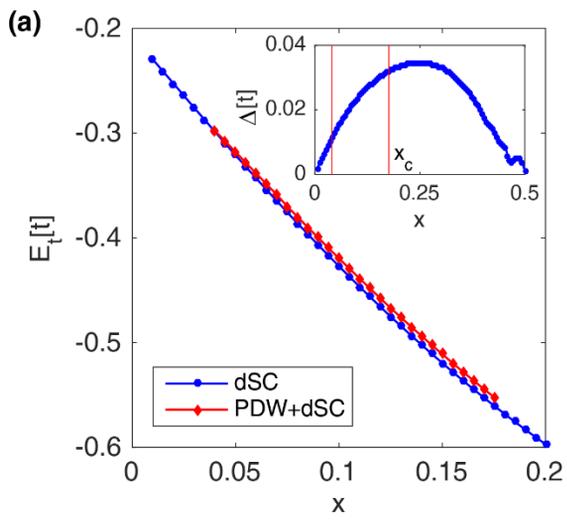

**(b)**

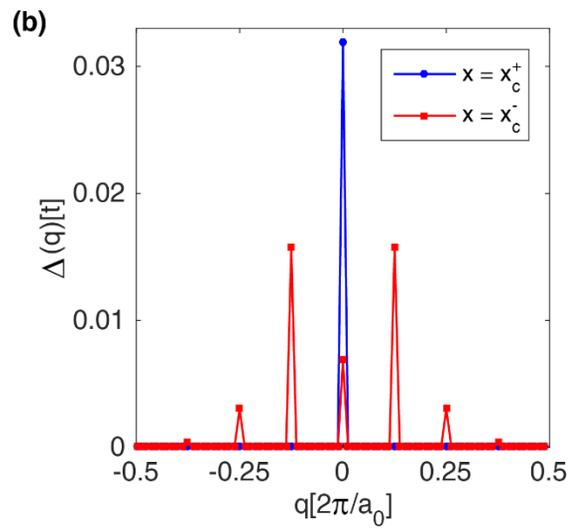

**Figure S2**

**(a)**
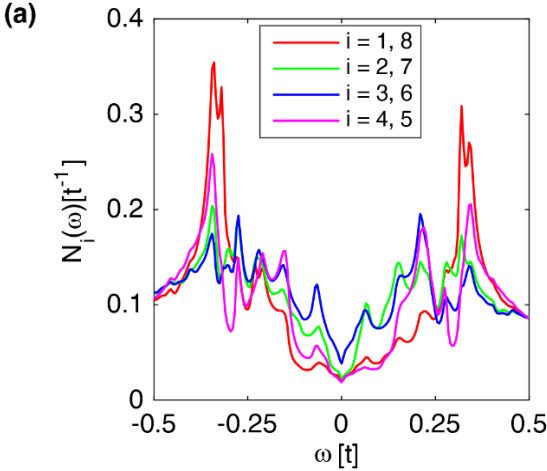

**(b)**
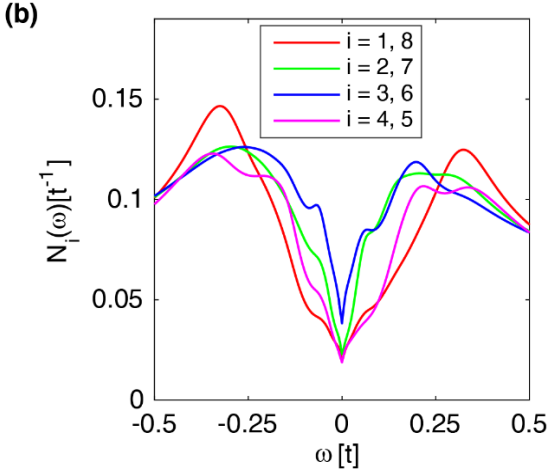

**Figure S3**

**(a)**

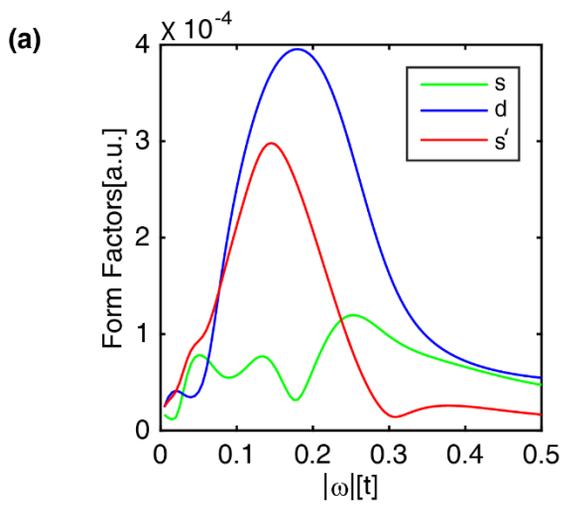

**(b)**

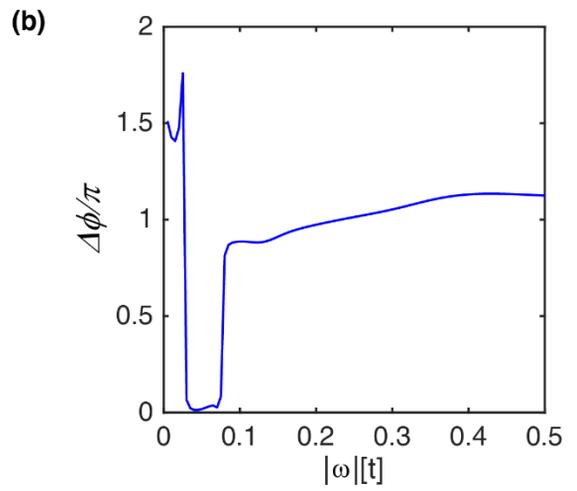

**Figure S4**

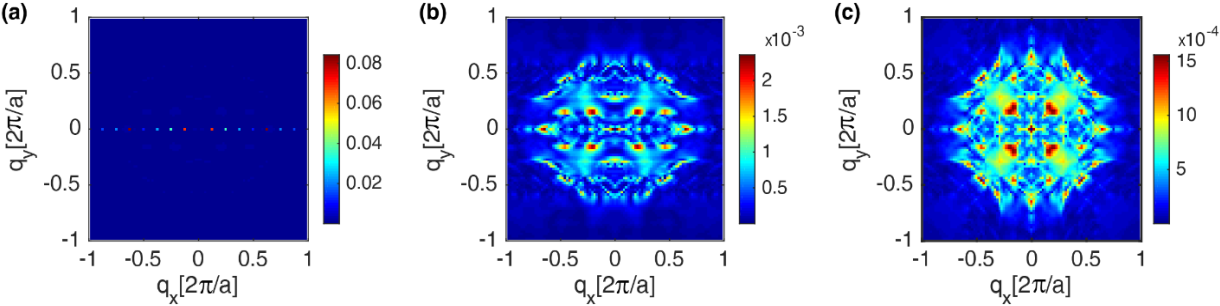

**Figure S5**

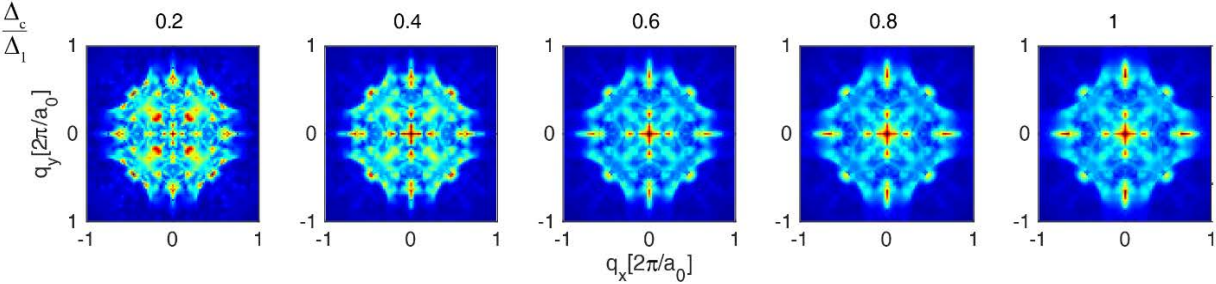

**Figure S6**

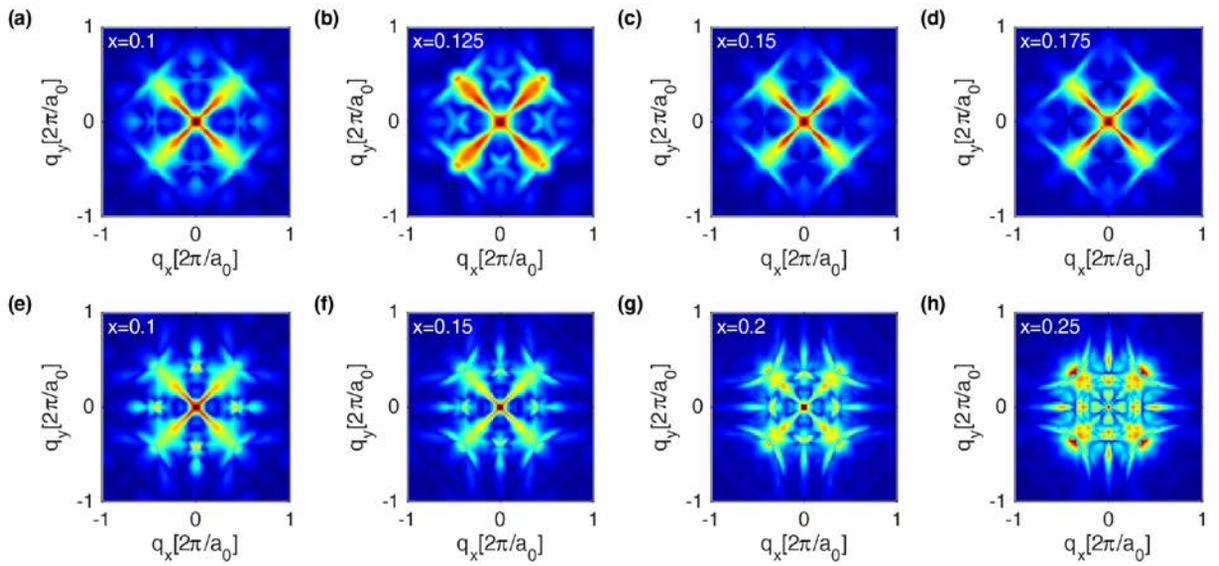

**Figure S7**

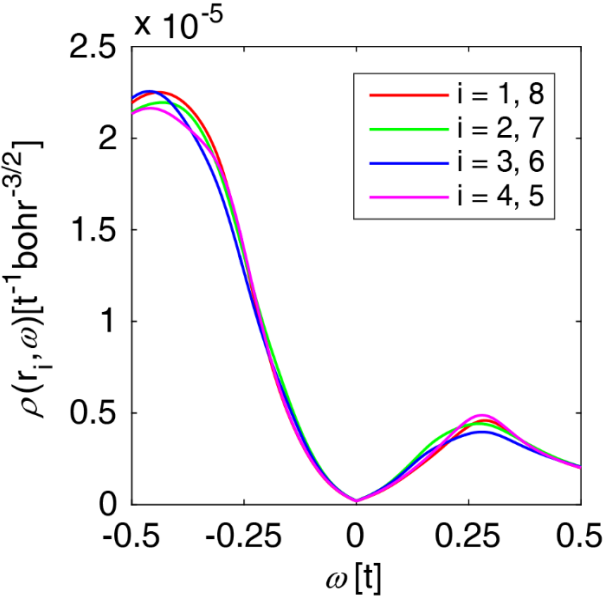

# Figure S8

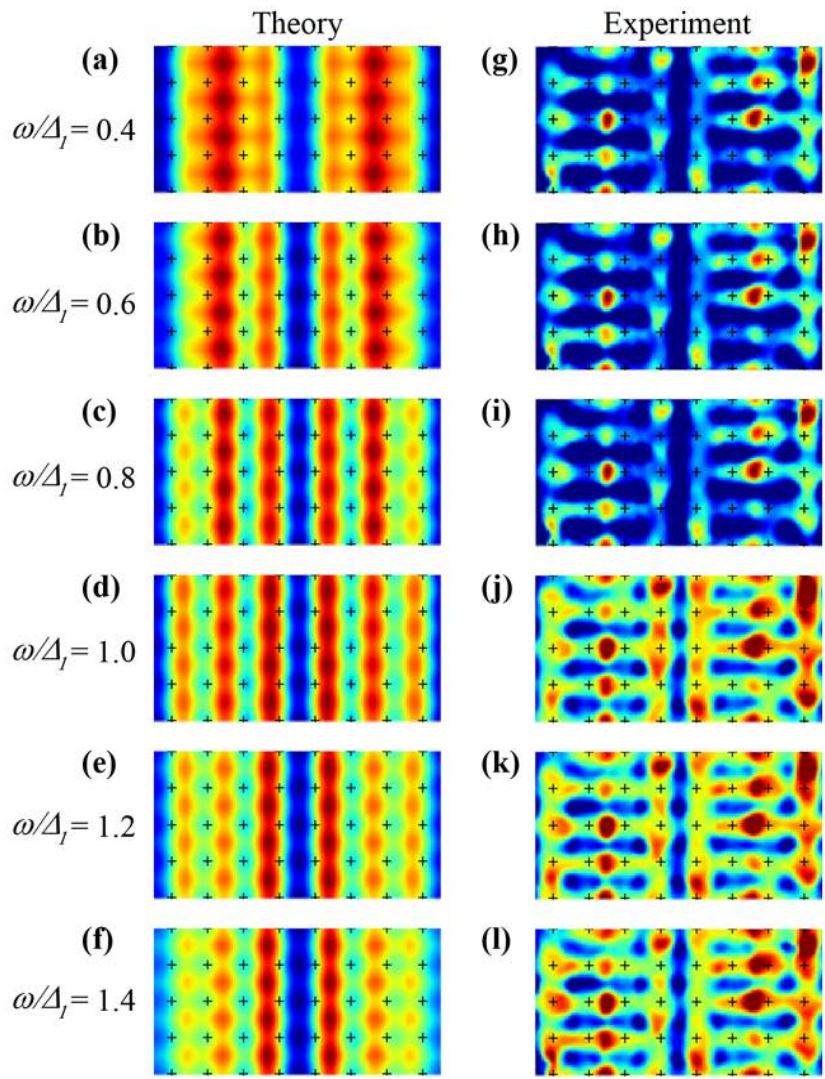

Theory       Experiment

**(a)** $\omega/\Delta_1 = 0.4$   **(g)**

**(b)** $\omega/\Delta_1 = 0.6$   **(h)**

**(c)** $\omega/\Delta_1 = 0.8$   **(i)**

**(d)** $\omega/\Delta_1 = 1.0$   **(j)**

**(e)** $\omega/\Delta_1 = 1.2$   **(k)**

**(f)** $\omega/\Delta_1 = 1.4$   **(l)**

Low ▬▬ High

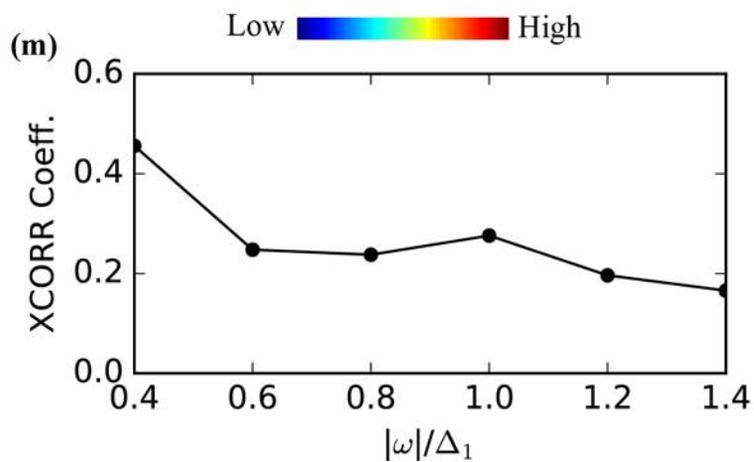

**(m)**

**Figure S9**

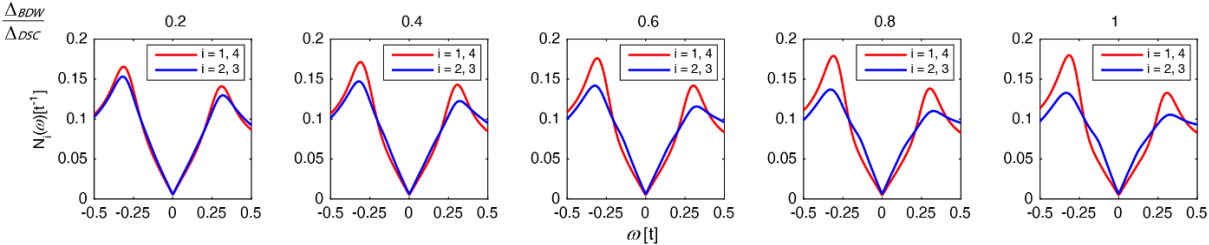

**Figure S10**

Theory             Experiment

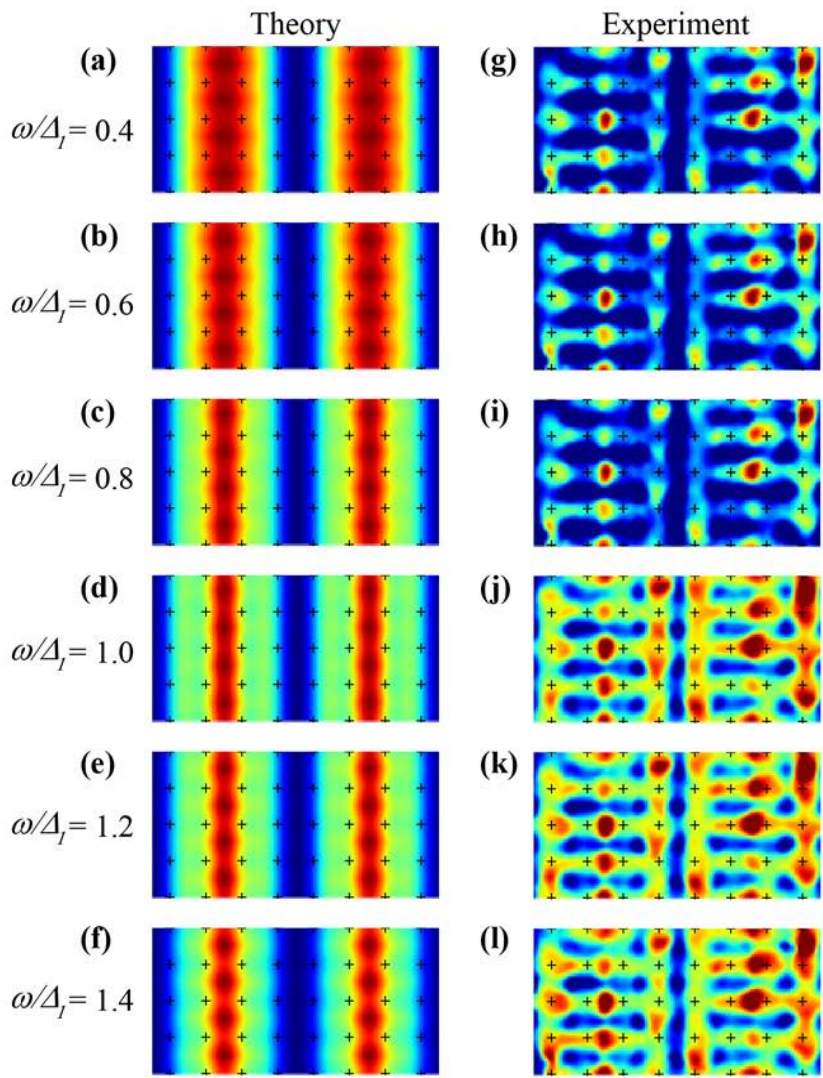

(a) $\omega/\Delta_1 = 0.4$    (g)
(b) $\omega/\Delta_1 = 0.6$    (h)
(c) $\omega/\Delta_1 = 0.8$    (i)
(d) $\omega/\Delta_1 = 1.0$    (j)
(e) $\omega/\Delta_1 = 1.2$    (k)
(f) $\omega/\Delta_1 = 1.4$    (l)

Low ▬▬▬ High

(m)

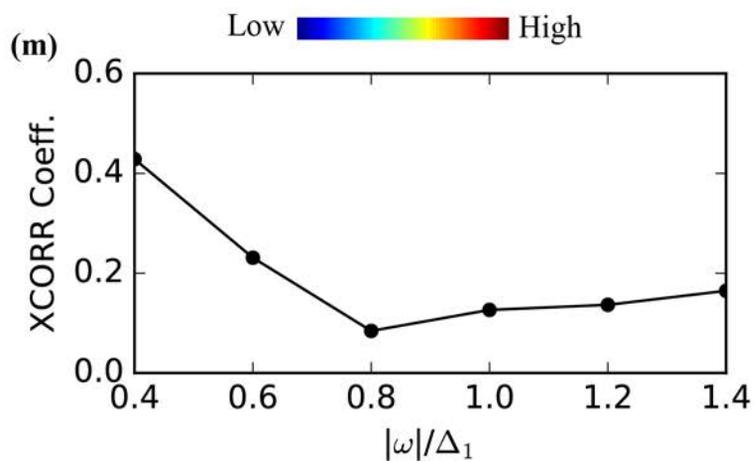

**Figure S11**

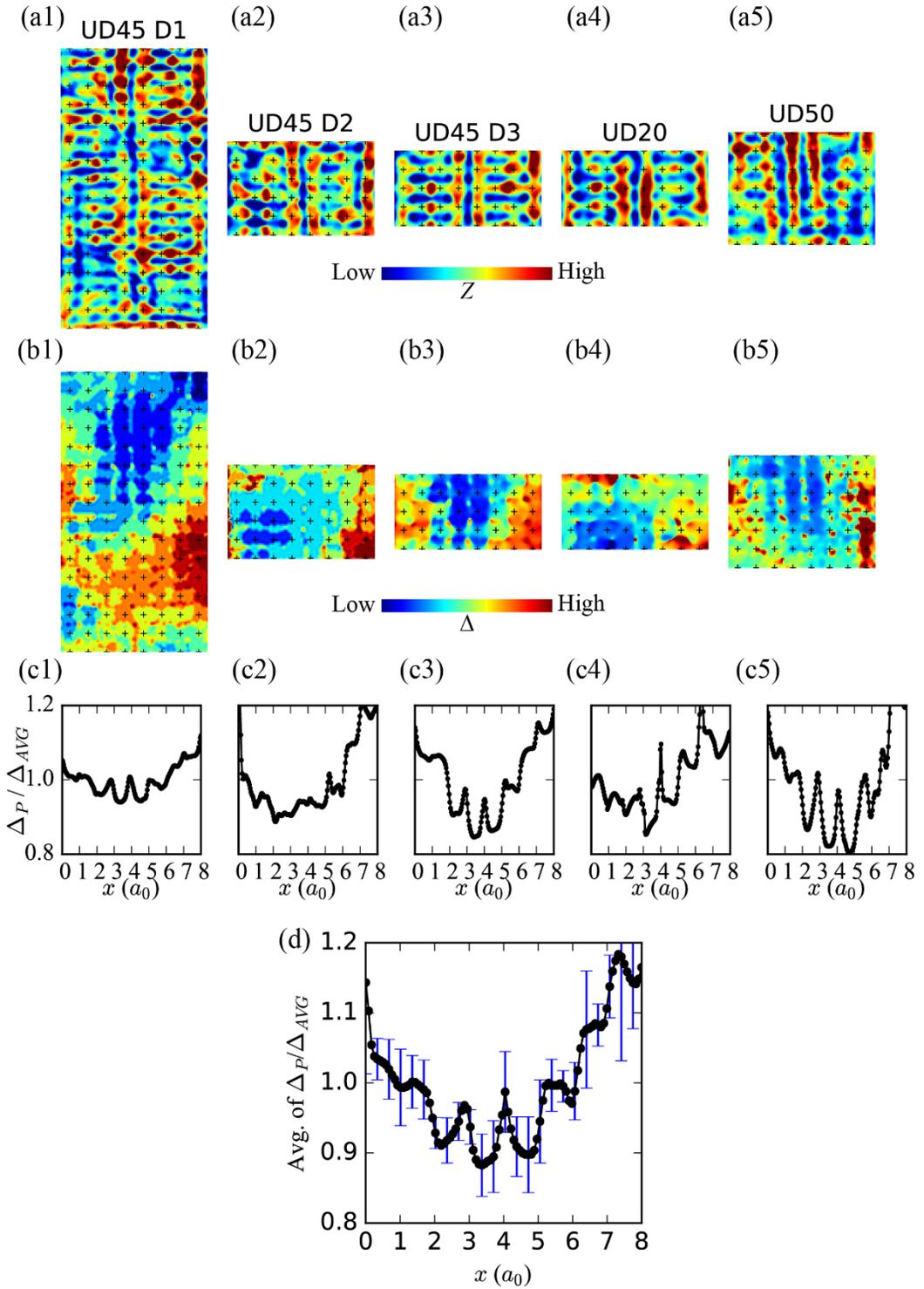